\documentclass[manuscript,screen]{acmart}
\AtBeginDocument{%
  }

\setcopyright{acmlicensed}
\copyrightyear{2025}
\acmYear{2025}
\acmDOI{XXXXXXX.XXXXXXX}

\acmConference[Conference acronym 'XX]{Make sure to enter the correct
  conference title from your rights confirmation emai}{June 03--05,
  2018}{Woodstock, NY}
\acmISBN{978-1-4503-XXXX-X/18/06}

\usepackage{tabulary}
\usepackage{subcaption}
\usepackage{multirow}
\usepackage{svg}

\begin{document}

\title[How Experienced Healthcare Professionals Work With Decision Support Systems in Clinical Practice]{``It Feels Weird to Just Rely on AI'': How Experienced Healthcare Professionals Work With Decision Support Systems in Clinical Practice}

\author{Simon W.S. Fischer}
\email{simon.fischer@donders.ru.nl}
\orcid{0000-0003-2992-6563}
\affiliation{%
  \institution{Donders Institute for Brain, Cognition, and Behaviour, Radboud University}
  \department{Dpt. of Human-Centred Intelligent Systems}
  \city{Nijmegen}
  \country{The Netherlands}
}

\author{Hanna Schraffenberger}
\email{hanna.schraffenberger@ru.nl}
\orcid{0000-0003-1847-2754}
\affiliation{%
  \institution{Interdisciplinary Hub for Digitalization and Society (iHub) and Institute for Computing and Information Sciences (iCIS), Radboud University}
  \city{Nijmegen}
  \country{The Netherlands}
}
\author{Miranda L. van Hooff}
\orcid{0000-0001-5313-6436}
\affiliation{%
  \institution{Radboud UMC}
  \department{Dpt. of Orthopaedic Surgery}
  \city{Nijmegen}
  \country{The Netherlands}
}
\affiliation{%
  \institution{Sint Maartenskliniek}
  \city{Nijmegen}
  \country{The Netherlands}
}

\author{Casper Reijnen}
\orcid{0000-0001-6873-7832}
\affiliation{
    \institution{Radboud UMC}
    \city{Nijmegen}
    \country{The Netherlands}
}

\author{Serge Thill}
\email{serge.thill@donders.ru.nl}
\orcid{0000-0003-1177-4119}
\affiliation{%
  \institution{Donders Institute for Brain, Cognition, and Behaviour, Radboud University}
  \department{Dpt. of Human-Centred Intelligent Systems}
  \city{Nijmegen}
  \country{The Netherlands}
}

\author{Pim Haselager}
\email{pim.haselager@donders.ru.nl}
\orcid{0000-0002-4077-9560}
\affiliation{%
  \institution{Donders Institute for Brain, Cognition, and Behaviour, Radboud University}
  \department{Dpt. of Human-Centred Intelligent Systems}
  \city{Nijmegen}
  \country{The Netherlands}
}

\renewcommand{\shortauthors}{Fischer et al.}

\begin{abstract}
  Decision support systems (DSS) assist with clinical decision-making. Little is known, however, about the decision-making process with DSS in clinical practice and the extent to which clinicians consider or rely on the predictions. Studies on human-AI decision-making usually focus on laypeople, simplified decision tasks, or one-shot interactions with prototypes that are not integrated into existing workflows. To gain insights from real-world practice, we conducted an online survey ($n=12$) and semi-structured interviews ($n=10$) with clinicians from two different medical cohorts each working with a DSS. Findings show, amongst other things, that the interaction with DSS changes as experience increases, the predictions become more important in complex cases, that decisions are discussed with colleagues, and that workflow integration affects decision-making. Our findings on human, technical, contextual, and organisational factors contribute to a socio-technical perspective on DSS reliance behaviour. We  conclude by discussing the implications of our findings for the design of DSS.
\end{abstract}


\received{\today}

\maketitle

\section{Introduction}
\label{sect:intro}

Decision support systems (DSS) are adopted in clinical practice to assist healthcare professionals in making decisions regarding diagnosis or treatment selection. During the interaction with a DSS it is important that clinicians remain cognitively engaged and rely appropriately on the DSS advice. Neither should clinicians accept incorrect advice, nor reject correct advice. Thus, both overreliance and underreliance should be avoided or minimised \cite{Cabitza2026a}. Appropriate reliance can help to improve patient well-being, and is even required by regulations such as the European Artificial Intelligence Act within the framework of human oversight \cite{EuropeanDataProtectionSupervisor2025}.

Various approaches for stimulating cognitive engagement and thus supporting appropriate reliance have been proposed \cite{Bucinca2021,Tian2026,Bach2023,Li2026a}. Meta-studies on human-AI decision-making, however, identified a misalignment between tested systems and the application reality \cite{Lai2023a,Eckhardt2026,Raees2026b}. Most approaches focus on simplified decision-tasks, are evaluated with laypeople, and are not tested in real-world workflows \cite{Raees2026b}. Furthermore, many studies adopt a formal definition of appropriate reliance based on statistical decision-theory and quantitative metrics, thereby merely focusing on the decision outcome \cite{Schemmer2023,Guo2024}. This focus overlooks the decision-making process and the extent to which the decision-maker reflects on the DSS advice and weighs it against other information, which is particularly relevant in multifaceted, open-ended decision tasks that have no clear ground truth \cite{Matsumoto2025}. 
In view of this, too little is known about the decision-making process with DSS in daily (clinical) practice.

To gain insights from real-world contexts, we investigate how clinicians work with a DSS in clinical practice and, in particular, whether they have already formed a judgement before seeing the predictions, to what extent they reflect on the DSS predictions, and how they proceed if their judgement differs from the prediction. Our research question (RQ) is: 
\textit{How do clinicians work with a decision support system to make decisions in clinical practice?}
Based on this RQ, we have the following sub-questions: Is overreliance considered a problem by clinicians? Are there behavioural reliance patterns, and if so, what are they? What are the clinicians' information needs in order to improve their decision-making?
To answer these questions, we focus on two medical contexts each of which utilises a DSS, and both of which are based in the Netherlands. In the first cohort, clinicians work with a DSS to make decisions regarding the treatment of patients with chronic low back pain. In the second cohort, clinicians use a DSS to help determine the extent of lymph node metastases in patients with endometrial cancer and to make treatment decisions accordingly.

We conducted an online survey and expert interviews with a total of 12 healthcare professionals. We triangulate our data by combining the two data sources, survey and interviews, which originate from the two different, yet homogenous participant groups (i.e., clinicians working with a DSS). The interviews ($n_{cohort1}=7$; $n_{cohort2}=3$) provide detailed insights that help to contextualise the quantitative data ($n_{cohort1}=8$; $n_{cohort2}=4$), and together they paint a nuanced picture of the interaction and thus reliance behaviour on DSS.
Hence, this paper makes the following contributions: 1) We provide insights into how clinicians interact with a DSS when making decisions in clinical practice. In particular, we identify factors shaping the decision-making process relating to a) \textit{the human}, such as task familiarity, to b) \textit{the technology} or the DSS, such as explanations, to c) \textit{the contextual}, such as patient preferences, and to d) \textit{the organisation}, such as workflow integration. We therefore contribute to a socio-technical perspective on reliance behaviour \cite{Eckhardt2026}. 2) Based on our findings, we discuss pathways for the design of DSS and for approaches promoting appropriate reliance.

\section{Background and Related Work}
\label{sect:background}

We discuss relevant literature related to appropriate reliance, as well as interaction methods that focus on the decision-making process rather than merely on the outcome, and thus differ from the traditional \textit{recommendation-centric} approach.

\subsection{Appropriate Reliance in Human-AI Decision-Making}

For optimal decision outcomes in human-AI decision-making, it is important that the decision-maker appropriately relies on the DSS, thereby avoiding both underreliance and overreliance \cite{Raees2026b,Cabitza2026a}.
Appropriate reliance is often defined as the decision-maker's ability to reject incorrect DSS advice, and to accept correct DSS advice \cite{Raees2026b, Schemmer2023}. In the case of underreliance, the decision-maker rejects the correct DSS advice and instead follows their own (incorrect) judgement (i.e., incorrect self-reliance \cite{Schemmer2023}). This may be because of a high level of self-confidence, a desire for agency, or ignorance about the DSS performance \cite{Sunstein2025}. As a study suggests, underreliance occurs frequently \cite{Cabitza2026a}. In the case of overreliance, the decision-maker wrongly accepts the incorrect recommendation of the DSS, without following their own (correct) judgement (i.e., incorrect AI-reliance \cite{Schemmer2023}). Overreliance can happen for various reasons, such as low expertise regarding the task and thus low self-confidence, low AI literacy, the belief in the system's superiority (i.e., automation bias), time constraints, or low attention \cite{Passi2022}.
Rather than solely focusing on whether the system's advice was accepted or rejected, however, a definition of appropriate reliance should also take the decision-making process into account, e.g., to what extent information was evaluated or alternatives were considered.

Various studies have investigated reliance behaviour, yet meta-studies have raised concerns regarding their external validity for a number of different interconnected reasons \cite{Raees2026b,Eckhardt2026,Lai2023a}. 
First, many studies on human-AI decision-making focus on simplified or generic decision tasks, such as maze-solving \cite{Lai2023a,Eckhardt2026}. The findings from these controlled settings may, however, not reflect real-world decision-making behaviour, particularly in complex, dynamic decision-making contexts, such as healthcare.
Second, and closely related to the decision task, studies primarily evaluate the decision outcome, i.e., whether the decision was correct or not, in terms of performance or decision accuracy \cite{Lai2023a,Eckhardt2026}. Many decisions, however, involve uncertainty (i.e., there is no clear ground truth), and are not (strictly) binary, meaning they cannot simply be categorised as correct or incorrect \cite{Sivaraman2025a}. In these cases, the formal definition of appropriate reliance is therefore not sufficient.
Third, laypeople or crowdworkers are invited for assessing reliance behaviour, even for domain-specific tasks \cite{Eckhardt2026}. Laypeople, however, lack relevant knowledge, which is why, compared with professionals or experts, they are likely to be less able and less confident in assessing the system's output appropriately \cite{Ford2023,Gaube2021}.
Lastly, several studies use novel `pretend' prototypes (e.g., \textit{Wizard of Oz} studies) in individual interaction sessions \cite{Eckhardt2026}. These systems might, however, not resemble or fit into existing decision workflows \cite{Lai2023a,VanDort2026}. Consequently, during these single interactions, decision-makers might show different reliance patterns than they would during prolonged use \cite{Dratsch2023}.

While various studies on human-AI decision-making and reliance behaviour provide relevant insights that can inform the design of potential DSS, they should nevertheless be complemented with insights drawn from real-world contexts. So, in addition to psychological or human factors \cite{Kuper2025,Dogru2025,DeSouzaFilho2026}, such as the confidence of the decision-maker, which is affected by their expertise (e.g., laypeople vs. experts) and DSS familiarity (e.g., single interactions with novel systems vs. prolonged use), organisational factors and existing workflows, in which the DSS is embedded and in which the decision-maker may receive support from colleagues (e.g., second opinion), should also be taken into account. Moreover, many decision-making contexts are multifaceted, consisting of multiple sources of information and successive decision-making steps, where different options and consequences need to be weighed up \cite{Matsumoto2025, Smith2008, Bate2012}. The clinical decision-making process involves, for example, shared decision-making with the patient, taking into account their preferences or life circumstances, as well as the possible presence of multiple medical conditions (i.e., comorbidities).

Against this backdrop, \citet{Eckhardt2026} advocate a socio-technical perspective on AI reliance, in which the interaction between \textit{the social component} (e.g., human decision-maker) and \textit{the technical component} (e.g., DSS) is considered within a specific context or \textit{environment} (e.g., the organisation or workplace). 
In this paper, we take such a perspective by investigating how experienced decision-makers integrate DSS predictions into their reasoning in clinical practice. %
In doing so, we understand appropriate reliance as the process of evaluating, reflecting, and cognitively engaging with the DSS advice. Others have described this as a form of ``negotiation'' \cite{Sivaraman2023}. 
We view human-AI decision-making as intertwined and entangled, in which sense-making arises during the interaction and interplay of the socio-technical components, namely the human, technical, and environment \cite{Fischer2024a,Frauenberger2020}. We therefore focus more on the decision-making \textit{process} than on the outcome and whether in/correct advice was accepted or rejected \cite{Zhang2024a}, thereby extending the common definition of appropriate reliance.

\subsection{Supporting the Decision-Making Process}

According to \citet{Miller2023}, most DSS or AI systems follow a \textit{recommend and defend} approach, in which a \textit{recommendation} or prediction is provided based on some input data, which is then often \textit{defended} with an accompanying explanation. Explanations, however, rarely lead to appropriate reliance as they can increase the trustworthiness of incorrect DSS advice \cite{Fok2024}. \citet{Miller2023} therefore suggests that DSS should help the decision-maker to evaluate different hypotheses or options by providing evidence for and against them without making a recommendation \cite{Cabitza2025,Zhang2024a}. 
Similarly, different interaction methods, so-called cognitive forcing functions or frictional design, have been proposed to encourage decision-makers to reflect and think critically about the DSS predictions or recommendations \cite{Bucinca2021,Fogliato2022,Bach2023,Cabitza2024b,deJong2025}.
These interaction methods follow a so-called \textit{secondary assistance pattern} \cite{Gomez2025}, in which the DSS does not (only) provide a prediction, but additional information such as reflective questions \cite{Tian2026,Li2026a} or feedback \cite{Reicherts2025} that facilitate the decision-making \textit{process}. 

In other studies, clinicians reported that they have the expertise to make decisions and would thus rather prefer support in intermediate stages of the decision-making process \cite{Liao2020,Bach2023,Zhang2024}, such as investigating alternative outcomes or hypotheses through counterfactual reasoning or what-if analysis \cite{Wexler2019}.
To potentially provide secondary assistance, one sub-question of our research focuses on the information needs of clinicians with regard to improving their decision-making. To this end, we use a taxonomy that identifies ten aspects at play in human-AI decision-making, such as data set, the causal structure of the recommendation, or alternatives to a prediction \cite{Fischer2025a}. 
In our survey, we ask clinicians which of these aspects they consider and whether they would like more information or support regarding those aspects during decision-making. For example, we ask participants to what extent they consider alternatives during decision-making, and to what extent they would like to know more about such alternative outcomes.
The results contribute to a better understanding of the decision-making process, and can inform the design of DSS or secondary assistance, which aim to promote appropriate reliance.

\section{Clinical Use Cases}
\label{sect:use-case}

We focus on two real-world clinical use cases, in each of which clinicians work with a DSS to make decisions regarding the treatment of 1) chronic low back pain and 2) lymph node metastases in endometrial cancer.

\subsection{Cohort 1: Treatment Prognosis for Chronic Low Back Pain}

\begin{figure}
    \centering
    \includegraphics[width=0.65\linewidth]{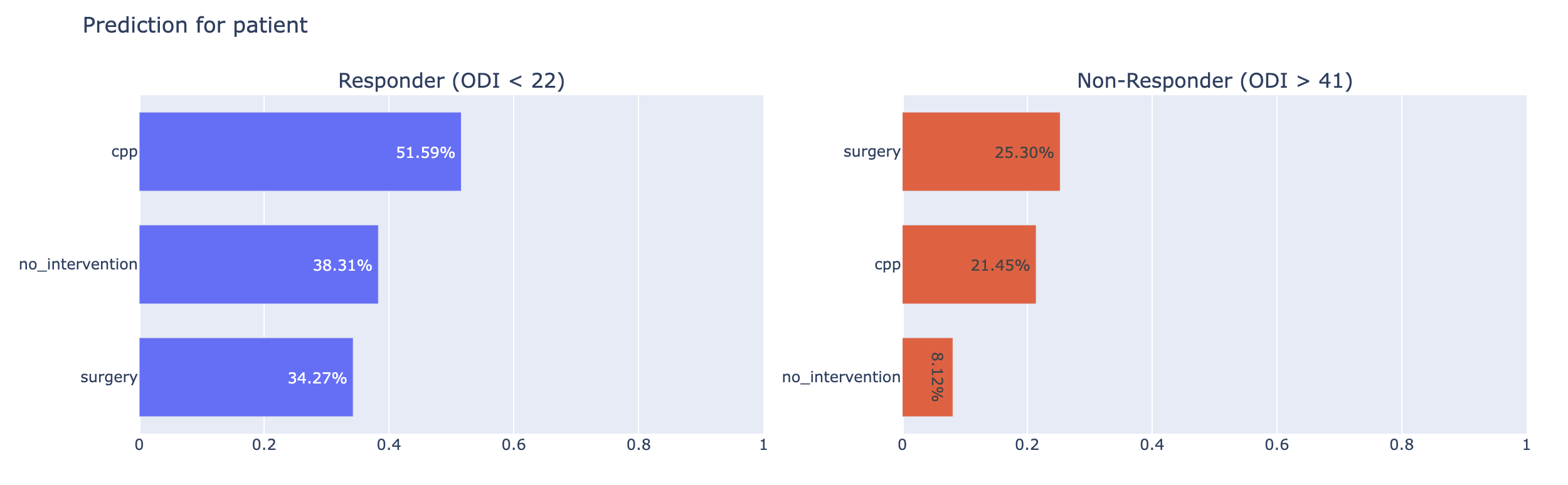}
    \caption{A replicated and anonymised visualisation of the DSS interface used in cohort 1, in English and without the hospital logo. The visualisation of the predictions resembles the actual DSS used in clinical practice. The DSS provides predictions for three treatment options with two outcomes each: on the left whether the patient is expected to \textit{respond} (i.e., success), and on the right whether they are expected to \textit{not respond} (i.e., failure) to the treatment.}
    \label{fig:ndt-interface}
    \Description{The image depicts an interface of a decision support system used in the medical cohort 1. The interface shows the predictions of six treatment options for chronic low back pain in the form of bar charts.}
\end{figure}

For our first use case, we collaborated with hospital clinicians at a spine department who use a DSS to determine the treatment for patients with chronic low back pain (CLBP). Low back pain is the leading cause of disability in most countries \cite{Ferreira2023}, and is a complex condition, the exact cause of which is often unknown \cite{Coupe2016}. In many cases, it is thus uncertain whether a treatment will lead to an improvement in the patient's functional status. Hence, there is no clear ground truth. 

To more effectively treat patients, researchers at the spine department defined predictive patient characteristics and developed clinical prediction rules that based on those characteristics estimate treatment outcomes (i.e., prognosis) \cite{Coupe2016}. The patient characteristics are gathered by a patient self-report questionnaire and take the form of tabular data. The prediction rules are implemented in the DSS as logistic regression functions. The DSS considers three treatment options: spine surgery, a combined physical and psychological program, and physiotherapy (Fig.~\ref{fig:ndt-interface}). For each treatment option, the DSS provides one prediction for expected `success' (i.e., responder) and one for `failure' (i.e., non-responder). Importantly, the success and failure rates are calculated independently with slightly different variables, resulting in six treatment prognoses (3 treatment options $x$ 2 outcomes). 
The DSS is deployed at only one hospital and is used by a total of 11 clinicians in the spine department. Most patients who come to the spine department have been suffering from back pain for several years and have already consulted other doctors and undergone previous treatments (e.g., surgery). The clinicians therefore deal with complex cases and need to view the DSS predictions critically. 

\subsection{Cohort 2: Risk Estimation of Lymph Node Metastasis in Endometrial Cancer}

\begin{figure}
    \centering
    \includegraphics[width=0.6\linewidth]{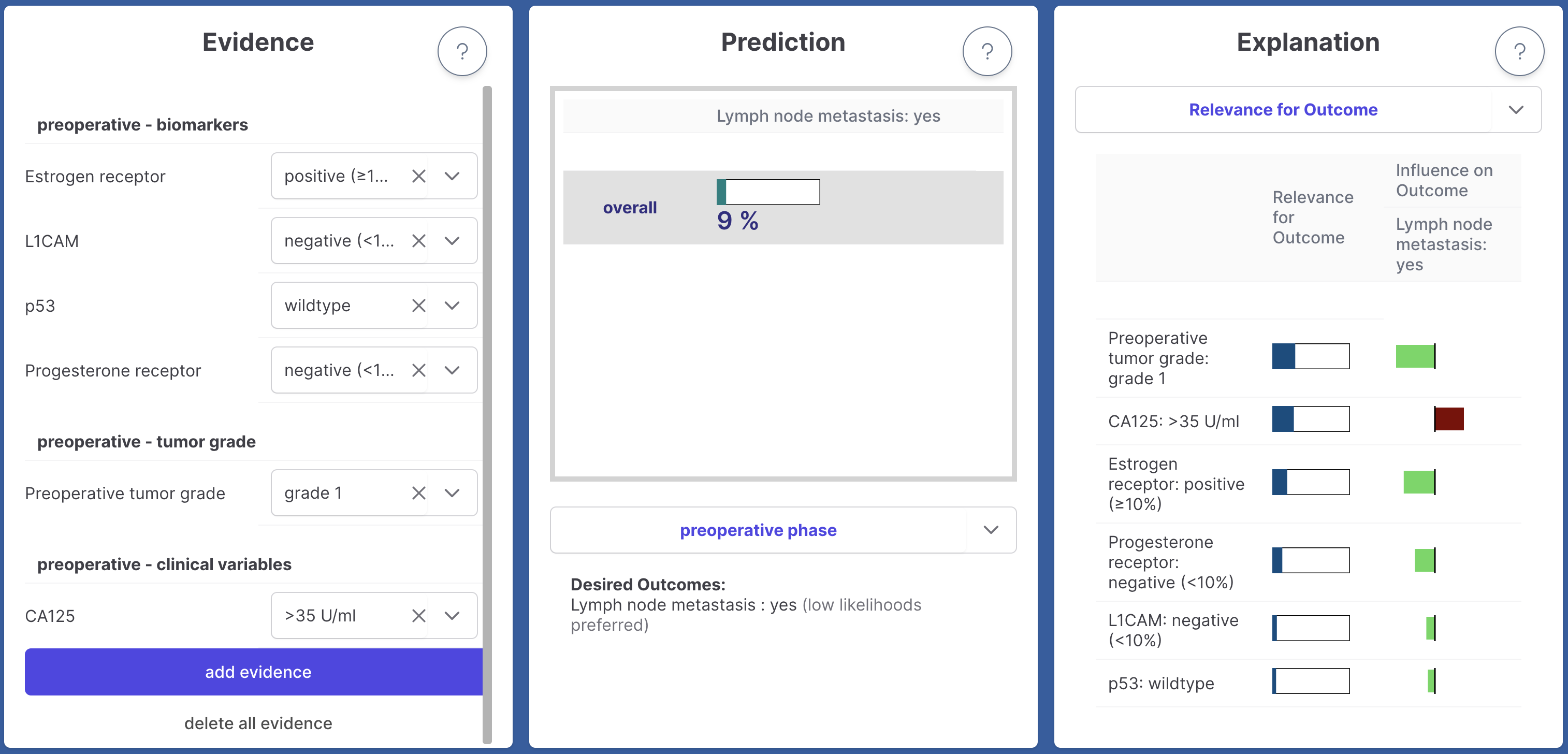}
    \caption{Interface of the DSS used in cohort 2. In the left column, the features or patient characteristics can be added. In the middle column, the risk score is shown. In the right column, an explanation of this prediction, in the form of feature contributions, is provided.}
    \label{fig:endo-interface}
    \Description{Interface of the DSS used in cohort 2. In the left column, the features or patient characteristics can be added. In the middle column, the risk score is shown. In the right column, an explanation of this prediction, in the form of feature contributions, is provided.}
\end{figure}

For our second use case, we collaborated with clinicians at various gynaecology departments. As part of a larger pilot study, a DSS has been introduced in 12 different hospitals at the beginning of 2026 and is currently being used by about 50 professionals in clinical practice. The DSS estimates the risk, of patients with endometrial cancer having lymph node metastases; a process also known as staging (e.g., stage 1, 2, 3). Endometrial cancer is a malignant condition of the uterus and the most common type of gynaecological cancer. There is a risk that the primary tumour spreads to the lymph nodes, and if the metastases are not treated early, the survival rate drops significantly \cite{Han2024}. So while the patient is already diagnosed with endometrial cancer, the clinicians use the provided risk score of lymph node metastases to make treatment decisions, whether to only remove the uterus or the lymph nodes as well. Lymph node removal is a major surgery with complications and only a small number of patients ($\sim$10\%) with endometrial cancer have metastases, therefore appropriate patient selection is important.

The DSS, implemented as a Bayesian network, combines multiple data sources in tabular form, such as biomarkers, blood tests and tumour grade. In addition to the estimated risk, the DSS displays an explanation for the prediction in the form of feature contributions (Fig.~\ref{fig:endo-interface}).

\subsection{Comparison of Both Decision Support Systems}

\begin{table}[t]
    \small
    \caption{Comparison of the two decision support systems used in Cohort 1 and Cohort 2.}
    \label{tab:dss-compare}
    \centering
    \begin{tabulary}{.7\textwidth}{p{1.3cm}p{1.3cm}p{2cm}p{1.4cm}p{2.2cm}p{1.8cm}p{1.8cm}}
    \toprule
         &
         \textbf{Use} &
         \textbf{Prediction} &
         \textbf{Model} & 
         \textbf{Data Source} &
         \textbf{Data Entry} &
         \textbf{Explanation}
         \\
    \midrule
         \textbf{Cohort 1} &
         Treatment Selection &
         Success/Failure of three treatments (\textit{Prognosis}) &
         Logistic\newline Regression &
         Patient Self-Report Questionnaire\newline (\textit{subjective}) &
         Automatic,\newline connected with patient record &
         None          
         \\
         \textbf{Cohort 2} &
         Treatment Selection &
         Risk estimation of metastases\newline (\textit{Staging}) &
         Bayesian Network &
         Biomarkers\newline (\textit{objective}) &
         Manual &
         Feature\newline Contribution         
         \\
    \bottomrule
    \end{tabulary}
\end{table}

Both DSS provide outcomes that help clinicians to decide on the type of treatment. Nevertheless, there are four main differences between the two DSS (Table~\ref{tab:dss-compare}), specifically with regard to 1) the prediction and treatment recommendation, 2) the data source, 3) data entry, and 4) explanations provided. 

The DSS in cohort 1, provides predictions or prognoses for six different treatment outcomes (Fig.~\ref{fig:ndt-interface}), thereby making a rather explicit treatment recommendation (1). The prognoses are based on a questionnaire, which asks patients, for example, about previous surgeries, pain levels, and expectations regarding recovery. The data source is thus subjective (2). The questionnaire is connected to the hospital's electronic patient record, and the responses including treatment predictions are shown to the clinicians once they open the patient record without having to enter any data (3). Lastly, the DSS does not provide any explanation for the predictions (4). 

The DSS in cohort 2, provides a single prediction for the estimated risk of lymph node metastases. The treatment recommendation is thus implicit (1): If the estimated risk is high, the recommendation is to remove the lymph nodes; if the estimated risk is low, the recommendation is not to remove the lymph nodes. The prediction is based on objective data sources (2), such as blood tests or imaging results. Once the tests have been performed or the imaging data has been analysed, the clinicians must enter the data manually into the DSS in order to view the prediction for that patient (3). %
Lastly, the DSS provides an explanation in the form of feature contributions to the prediction (4).

\section{Methods}
\label{sect:methods}

Our research, conducted at the two medical cohorts, consists of two parts, namely 1) an online survey, and 2) individual, semi-structured interviews.
We developed both methods as part of a larger study on human-AI decision-making in the clinical context. As part of a parallel research project, we evaluated a prototype intervention aimed at increasing cognitive engagement and thereby mitigating overreliance on the DSS predictions. To this end, we replicated the functionality of the DSS used in cohort 1 and conducted interviews to gather feedback on our prototype. In addition, we asked participants about the decision-making process with the DSS. 
In this paper, we analyse the interview data we collected in cohort 1 with a focus on the general decision-making process with the DSS, rather than focusing on the feedback regarding the specific prototype (which is the focus of another paper). Furthermore, we analyse the survey data in this paper, which we have not used in other work, and we collected additional qualitative and quantitative data from cohort 2 specifically for this research. The study was approved by the ethics committee of our institution.

\subsection{Participants}

We recruited participants through contacts with clinical researchers from both medical cohorts who were involved in the development of the respective DSS. They have forwarded our online survey and call for participation to the relevant clinicians. In the survey, respondents could provide their contact details to be contacted for further studies (i.e., the interviews). Moreover, in cohort 1, the clinical researcher introduced us to a clinician, who then provided us with contact details of their colleagues. Accordingly, we recruited additional participants through snowball sampling. 

In cohort 1, we were able to recruit eight out of the 11 (72.7\%) clinical spine specialists working with the same DSS for chronic low back pain.
For cohort 2, we could recruit four clinicians out of the approximately 50 clinicians working with the same DSS across different hospitals. Although a larger number of participants from cohort 2 would have been desirable, the total of 12 participants from a relatively homogenous study population (i.e., clinicians working with a DSS) was sufficient to reach information saturation and provide an answer to our RQ. Similarly, a meta-review on qualitative research reports that information saturation is generally reached with 9-17 participants \cite{Hennink2022}.

Among the 12 participants were six orthopaedic surgeons, three gynaecologists, one physician, one physician assistant, and one radiation oncologist. Overall, our participants have extensive experience, both in terms of their many years of professional experience and in terms of working with a DSS. The characteristics of our participants are shown in Table~\ref{table:participants}. Participants could join only one part of our research (i.e., survey or interview), yet most voluntarily participated in both. 12 participants completed the survey, out of which 10 agreed to be interviewed. Each participant was informed about the purpose of this research and how their data would be used and stored, and they provided their written consent. For joining the interviews, participants were compensated with €15/h.

\begin{table}[t]
    \small
    \caption{Characteristics of our participants, divided into cohort 1 (P1-P8) and cohort 2 (P9-P12). Frequency of use was measured on a six-item scale ranging from ``Once or twice a month'' to ``Several times a day''. For a visualisation of the participant characteristics, see Fig.~\ref{fig:participant-characteristics}.}
    \label{table:participants}
    \centering
    \begin{tabulary}{\linewidth}{Lp{2.8cm}p{1cm}p{1.4cm}p{2.5cm}LL}
    \toprule
         \textbf{ID} &
         \textbf{Profession} &
         \textbf{Exp. (y)} &
         \textbf{DSS Use (y)} &
         \textbf{Frequency of Use} &
         \textbf{Survey} &
         \textbf{Interview}
         \\
    \midrule 
         P1 &
         Orthopaedic Surgeon &
         9+ &
         6-8 &
         Several times a week &
         $\checkmark$ & 
         $\checkmark$
         \\
         P2 &
         Orthopaedic Surgeon &
         3-5 &
         3-5 &
         Several times a week &
         $\checkmark$ & 
         $\checkmark$
         \\
         P3 &
         Physician Assistant &
         6-8 &
         3-5 &
         Several times a week &
         $\checkmark$ & 
         $\checkmark$
         \\
         P4 &
         Orthopaedic Surgeon &
         9+ &
         9+ &
         Several times a month &
         $\checkmark$ & 
         $\checkmark$
         \\
         P5 &
         Orthopaedic Surgeon &
         9+ &
         6-8 &
         Once or twice a week &
         $\checkmark$ & 
         $\checkmark$
         \\
         P6 &
         Physician &
         9+ &
         9+ &
         Several times a week &
         $\checkmark$ & 
         $\checkmark$
         \\
         P7 &
         Orthopaedic Surgeon &
         6-8 &
         6-8 &
         Several times a week &
         $\checkmark$ & 
         $\checkmark$
         \\
         P8 &
         Orthopaedic Surgeon & 
         6-8 &
         6-8 &
         Several times a month &
         $\checkmark$ & 
         $x$
         \\
         \midrule
         P9 &
         Radiation Oncologist &
         9+ &
         3-5 &
         Once or twice a month &
         $\checkmark$ &
         $\checkmark$
         \\
         P10 &
         Gynaecologist &
         9+ &
         6-8 &
         Once or twice a week &
         $\checkmark$ &
         $\checkmark$
         \\
         P11 &
         Gynaecologist &
         9+ &
         6-8 &
         Several times a month &
         $\checkmark$ &
         $x$
         \\
         P12 &
         Gynaecologist &
         9+ &
         6-8 &
         Several times a month &
         $\checkmark$ &
         $\checkmark$
         \\
        \bottomrule
         &
         &
         &
         &
         &
         $n=12$ &
         $n=10$
         \\
    \end{tabulary}
\end{table}

\subsection{Online Survey}

We conducted an online survey in Qualtrics. Before distributing the survey, we presented an initial set of questions to a clinical researcher, who was involved in the development of the DSS from cohort 1 and is one of the authors of this paper. After incorporating the feedback, we presented a revised questionnaire to the clinician from cohort 1 who had been introduced to us, and who suggested minor changes. The final survey consists of six thematic sections, namely demographics, attitudes towards the DSS, confidence levels, interaction behaviour, considerations of various information, informational needs, and open questions (see supplementary material). 
On average, it took participants 13 minutes to complete the survey, including the introduction and consent section. We sent out the survey prior to the interviews.

Demographic data was collected using multiple-choice questions regarding age, gender, profession, years of professional experience, years of experience working with the DSS, primary use of the DSS, frequency of DSS use, and knowledge of artificial intelligence. For cohort 2, we slightly changed the question regarding years of experience working with the DSS given that the DSS had only been deployed beginning of 2026. We therefore asked about years of experience working with similar prognostic DSS in clinical practice.

To assess the attitudes towards the DSS, we used the Usefulness, Satisfaction, and Ease of use (USE) scale \cite{Lund2001}, which is a scale to measure subjective usability of a product or application \cite{Gao2018}. The USE scale consists of 30 items, divided into four dimensions, namely Usefulness, Ease of Use, Ease of Learning, and Satisfaction. Since we assume that a greater utility of the DSS also has greater impact on the decision-making process, we only used the eight items of the Usefulness dimension and two of the seven items in the Satisfaction dimension, which also indicate the usefulness (i.e., ``I am satisfied with it'', ``I would recommend it to a colleague'').
Accordingly, we used 10 out of the 30 items and excluded the dimensions Ease of Use (e.g., ``It is user friendly''), and Ease of Learning (e.g., ``I easily remember how to use it'') as they did not serve our purpose.

For the confidence of the decision-makers, we used a 5-point Likert scale (1: not at all confident, 5: very confident) assessing four confidence levels, namely confidence in their expertise, confidence in their decision \textit{before} using the DSS, confidence in their decision \textit{after} using the DSS, and confidence in their understanding of the DSS. We drew inspiration from the Decision Self-Efficacy Scale, which measures self-confidence in one's own decision-making ability \cite{OConnor1995}. 

To identify general interaction patterns with the DSS, we asked participants to rate, on a 5-point Likert scale (1: strongly disagree; 5: strongly agree), their level of agreement with the following 10 statements: 1-3) whether the prediction a) confirms, b) contradicts, or c) changes their judgement; 4) whether they consider the prediction; 5-6) whether they have formed a judgement a) before or b) after seeing the prediction; 7) whether they reflect on the predictions; 8) whether the prediction helps them reflect on their judgement; 9) whether they think colleagues prefer the predictions over their own judgement; and 10) whether trainees should practice without the DSS.

We then used 13 items that inquired about how often participants consider certain aspects of the DSS during decision-making. We derived these items from the above-mentioned taxonomy that identifies 10 elements relevant in human-AI decision-making \cite{Fischer2025a}.
For seven elements of the taxonomy, we derived one statement each, whilst for three elements we have two statements each, making a total of 13 items. The 13 survey items are as follows, with the taxonomy ID mapping given in brackets: data quality (Q1 case information); data reliability (Q1 case information); relevance of symptoms/factors for a specific prediction (Q2 relevance of data); which data the DSS considers and whether relevant data is missing (Q3 data set); how a prediction follows from the input data (Q4 causal structure of recommendation); alternatives to the strongest prediction (Q5 alternative to recommendation); own implicit assumptions (Q6 assumptions and expectations of decision-maker); patient's preferences (Q7 stakeholder preferences); consequences of the decision (Q8 consequences of decision); how the prediction changes if a certain input changes (Q9 change intervention); which change in symptoms could make a (desired) treatment more likely (Q9 change intervention); to what extent symptoms can change without affecting the prediction (Q10 model behaviour), and factors that can change the prediction (Q10 model behaviour and Q9 change intervention). 

We then repeated the same 13 items, this time phrased as whether the participants would like to know more about those aspects. In doing so, we derive potential informational needs during decision-making. 

At the end of the survey, we asked the following four open-ended questions: 1) What do you do if your judgement and the DSS prediction differ?; 2) What are the challenges you face in your decision-making process?; 3) How does the DSS help you with these challenges?; 4) What would you need to overcome these challenges and make better decisions (also beyond the DSS)? In addition, we provided an open-text field for any further remarks.

\subsection{Expert Interviews}

We conducted semi-structured interviews to gather more nuanced insights into how practitioners work with the DSS in daily practice. For cohort 1, all seven interviews took place in person at the practitioner's office. Six interviews lasted for 45-60 minutes each, while two lasted for around 30 minutes each due to the participants' time constraints. For cohort 2, one interview took place in person at the practitioner's office, lasting for about 50 minutes, and two interviews were conducted online via video-conferencing software, lasting about 30 minutes each. As mentioned, data collection from cohort 1 was part of a larger study in which we evaluated a prototype, hence the longer duration of the interviews compared to cohort 2. In cohort 2, the longer interview of approximately 50 minutes was due to a more elaborate introduction to the DSS. All interviews were conducted by the first author.

Our general interview procedure included questions about the participants' occupations, how long they have been using the DSS, whether they perceive the DSS useful, whether they regard overreliance as a problem, what they do when their own judgement differs from the DSS prediction, whether the explanations provided are helpful (cohort 2), and what would help them in their decision-making. 
We, however, made minor adjustments to our interview procedure for subsequent interviews where necessary by carrying out an initial, lightweight analysis before conducting the next one. This iterative approach allowed us to focus more closely on specific aspects in later interviews in order to gather further information. Accordingly, some interviews, for example, focused more on the integration of DSS into workflows, while others focused more on the importance of additional clinical information in the decision-making process.

\subsection{Data Analysis}

We audio-recorded the interviews with Audacity and transcribed them using noScribe \cite{Droge2026}. To analyse our interview data, we used thematic analysis, a flexible coding procedure \cite{Braun2006}.
The first author checked the transcripts for correctness and created the initial code book. Since we collected and analysed data iteratively, the process of creating and refining codes was also iterative. After each interview, we combined, deleted, and revised identified codes. 
Furthermore, we initially analysed the qualitative data separately from the quantitative data. Once we had identified (preliminary) themes, we refined and connected both thematically. So while we identified initial codes and themes primarily inductively and data-driven, without any pre-existing framework or codebook \cite{Braun2006}, the concepts we used to construct our survey, such as confidence, experience, or interaction behaviour, ultimately provided some guidance in the creation our final themes. We discussed the themes amongst the authors, and one clinical researcher from cohort 2, who is also a co-author of this paper, agreed with the themes identified.

We did not perform statistical tests on the quantitative data, given that our aim was not to compare the two cohorts, nor to find statistical relationships between certain factors. Besides, our sample size ($n=12$) is too small to provide meaningful statistical insights. Instead, we provide descriptive statistics on the survey responses for each cohort and use the quantitative data to triangulate the qualitative results.

\section{Results}
\label{sect:findings}

We present our results in response to our RQ. %
In total, we identified eleven themes: 1) Both familiarity with the task and with the DSS shape reliance behaviour; 2) DSS use changes with experience; 3) The same DSS is used for different purposes; 4) The DSS predictions make decision-making more concrete; 5) A critical attitude encourages clinicians to follow their own judgement; 6) The DSS predictions help clinicians reflect on their judgement; 7) The DSS predictions become more important when clinicians are in doubt; 8) Inexperienced clinicians may be more influenced by DSS predictions; 9) The DSS predictions are just one aspect of a multi-faceted decision-making process; 10) Decisions are discussed with colleagues; and 11) Workflow integration of the DSS affects decision-making.

\subsection{Both Familiarity With the Task and With the DSS Shape Reliance Behaviour}

\begin{figure}
  \centering
  \subcaptionbox{\label{fig:exp-prof}}[0.35\linewidth]{
    \includegraphics[width=\linewidth]{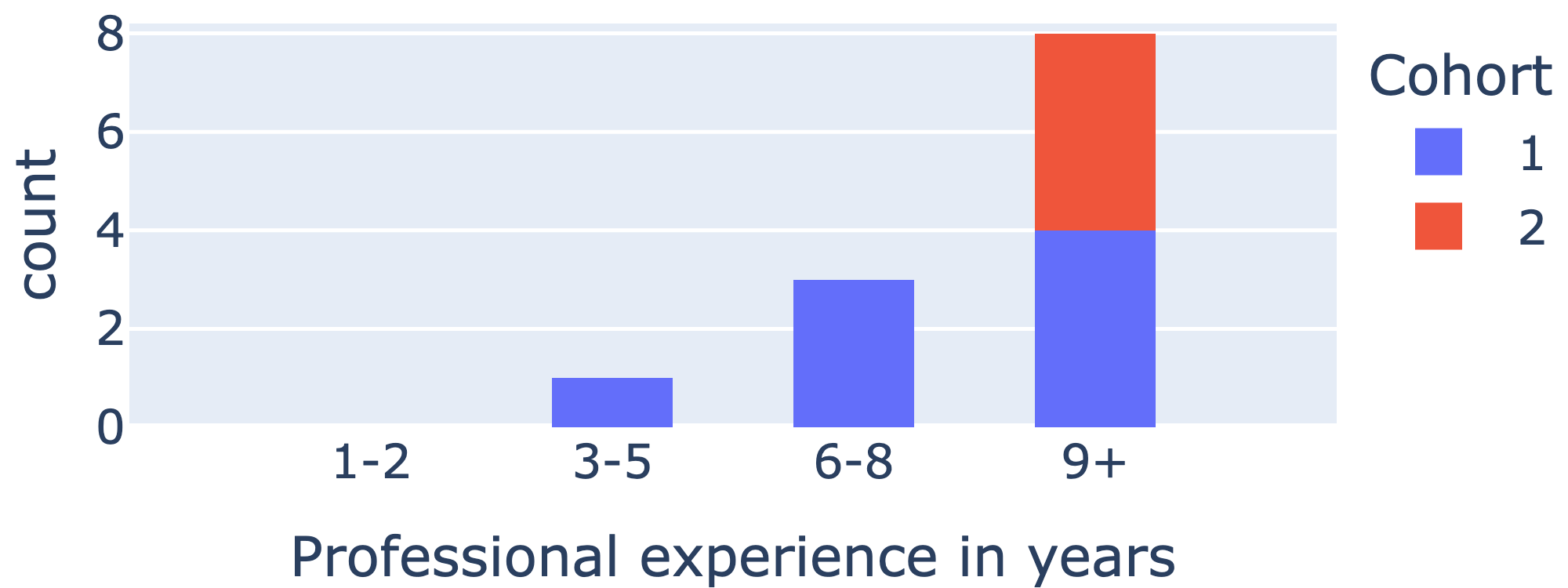}
  }
  \subcaptionbox{\label{fig:exp-dss}}[0.35\linewidth]{
    \includegraphics[width=\linewidth]{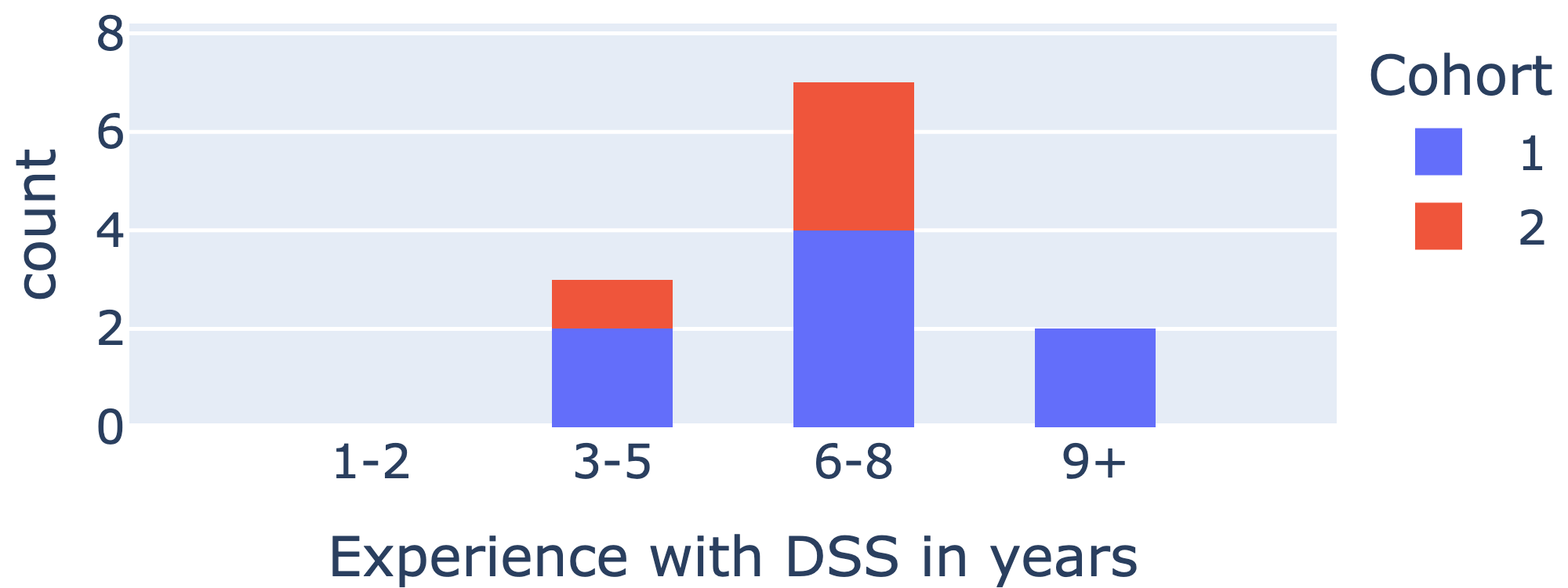}
  }
  \subcaptionbox{\label{fig:exp-pair}}[0.35\linewidth]{
    \includegraphics[width=\linewidth]{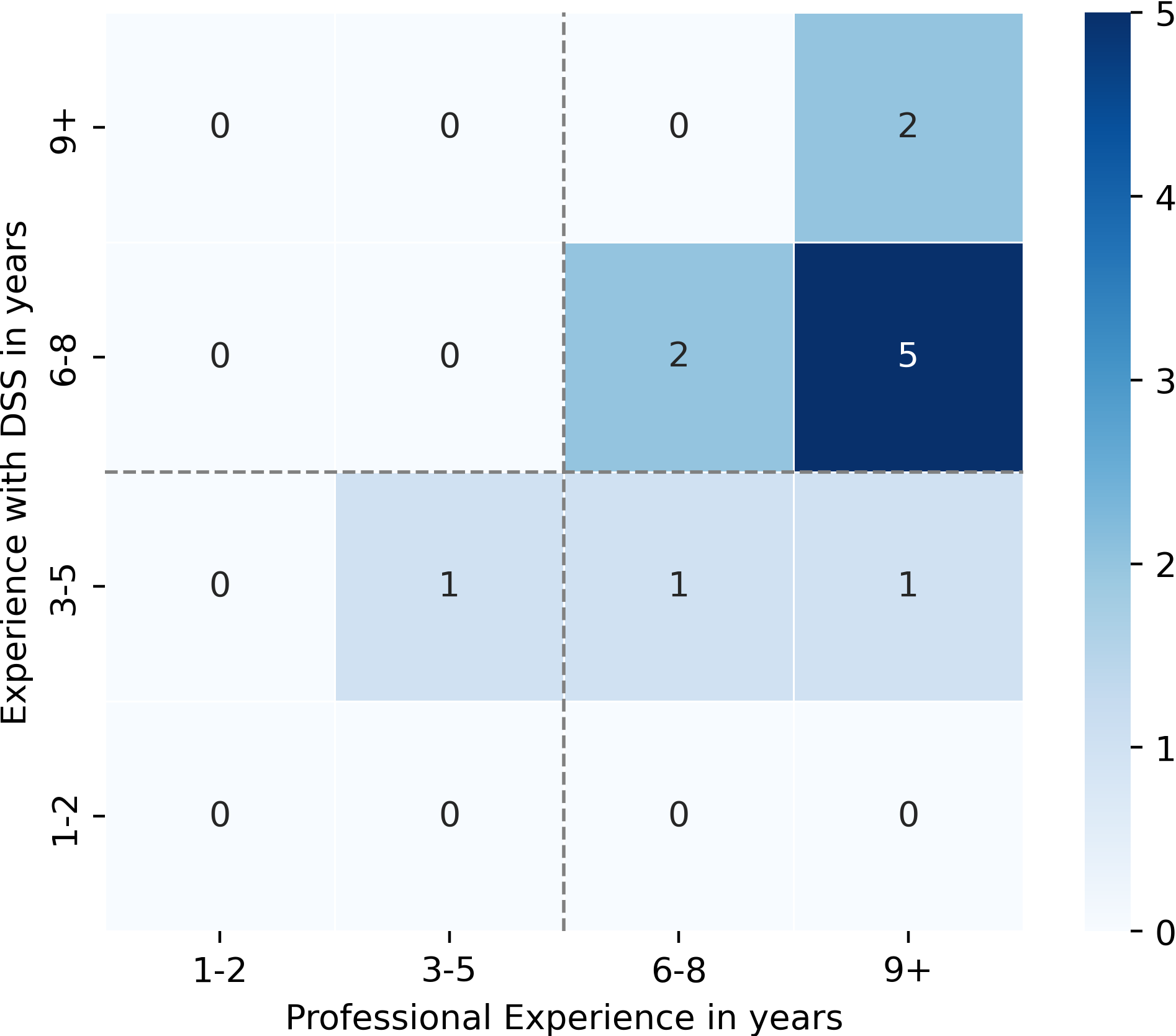}
  }
  \par
  \subcaptionbox{\label{fig:dss-freq}}[0.35\linewidth]{
    \includegraphics[width=\linewidth]{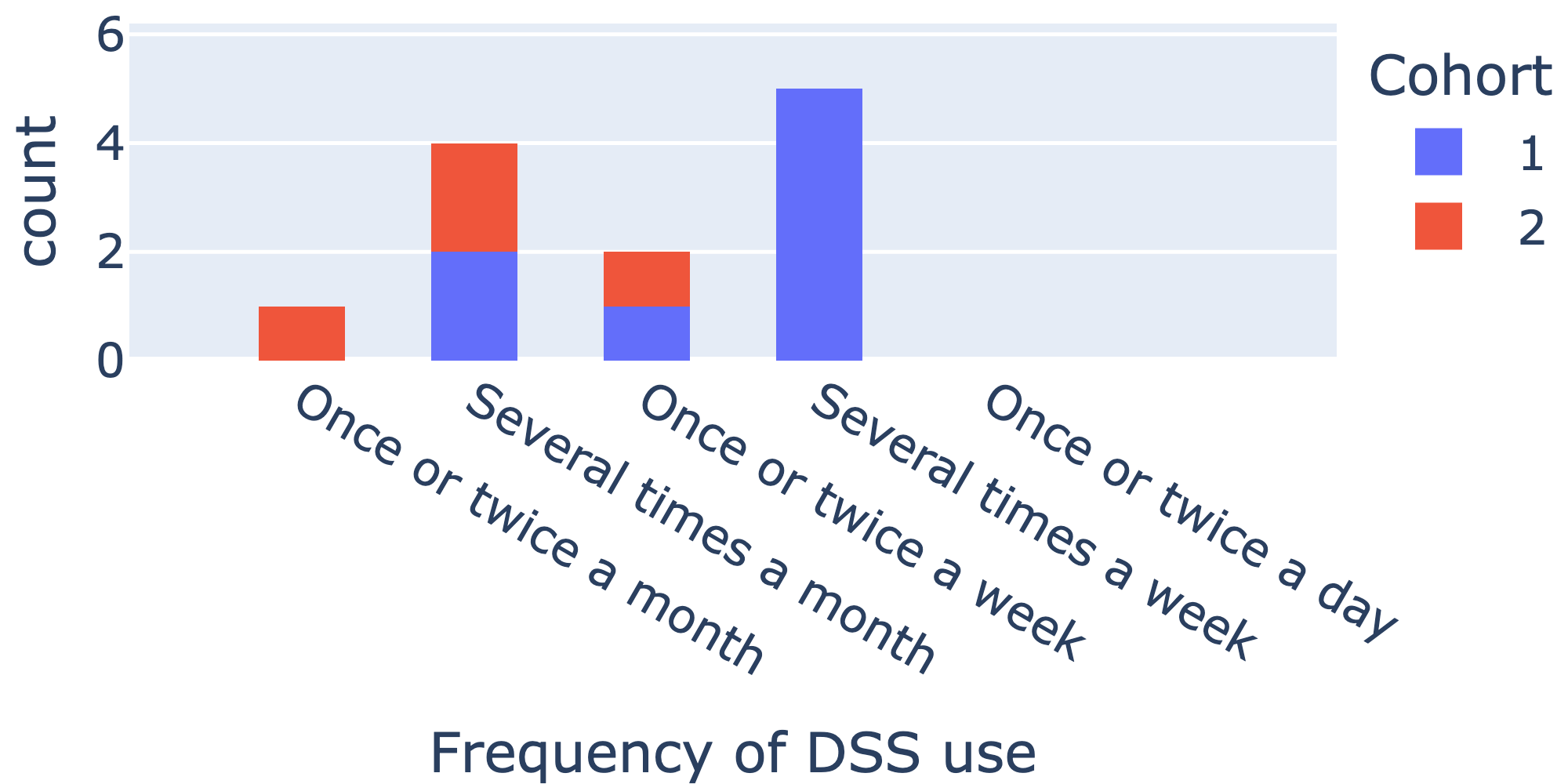}
  }
  \subcaptionbox{\label{fig:ai-knowledge}}[0.35\linewidth]{
    \includegraphics[width=\linewidth]{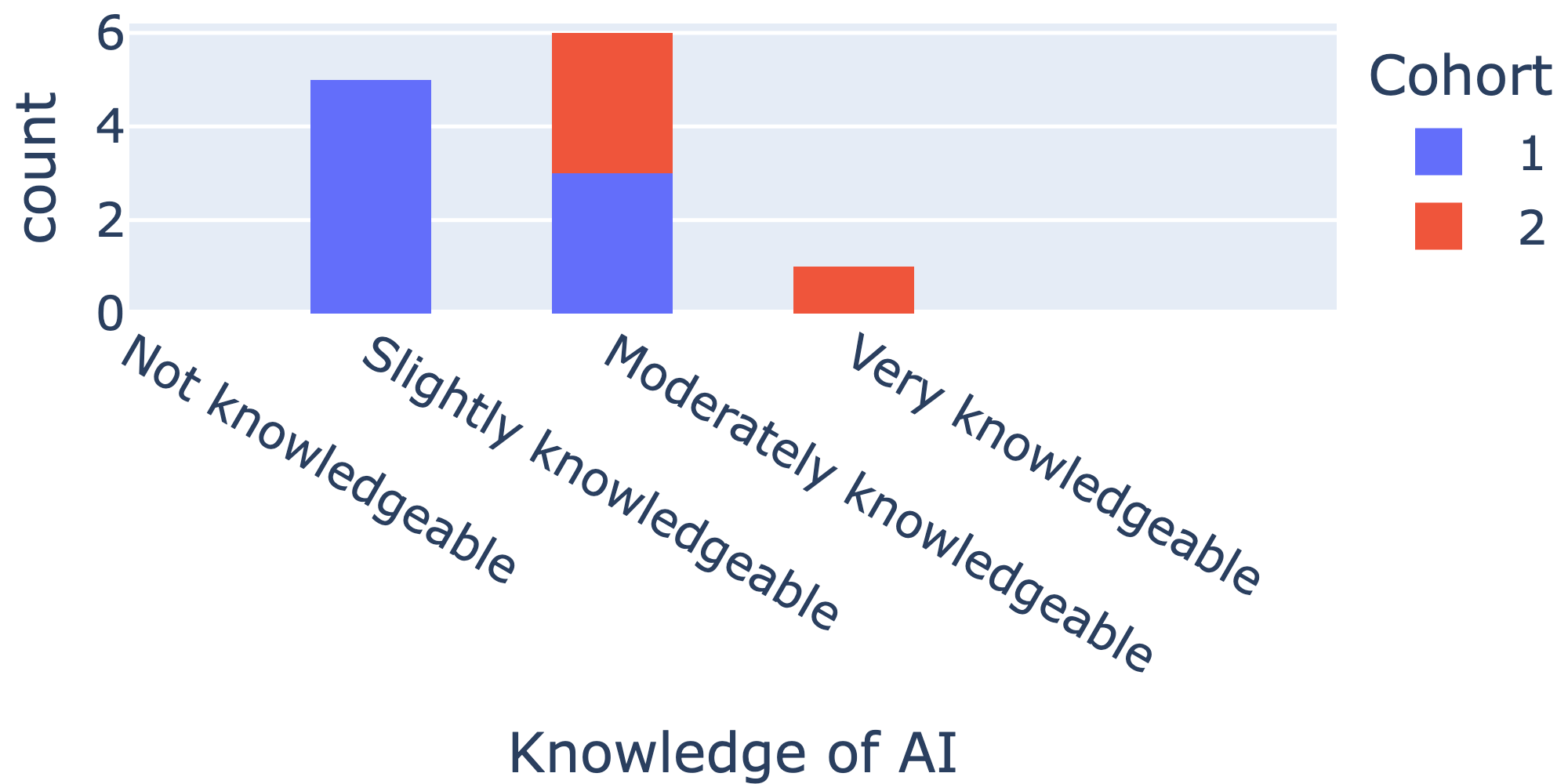}
  }
  \caption{Characteristics of our participants (see Table~\ref{table:participants}).}
  \label{fig:participant-characteristics}
  \Description{The characteristics of our participants from Table~\ref{table:participants} are represented as bar charts and as a heatmap, which shows the co-occurrence of experience levels for both professional experience and experience working with the DSS.}
\end{figure}

In general, our participants are very experienced. Out of the 12 participants, 11 have at least six years of professional experience (Fig.~\ref{fig:exp-prof}), with nine having six or more years' experience of using the DSS (Fig.~\ref{fig:exp-dss}). Thus, nine out of the 12 participants have the highest co-occurrence of experience levels, falling into the top right quadrant of Fig.~\ref{fig:exp-pair}. 
In addition, participants use the DSS rather frequently, from several times a month to several times a week (Fig.~\ref{fig:dss-freq}).
In the interviews, one participant mentioned that they ``always look at the DSS'' (P2) for each patient because they triage incoming patients.
Consequently, our participants are very familiar with both the decision-task, i.e., treatment selection, and with the use of the DSS (Fig.~\ref{fig:participant-characteristics}). 

\begin{figure}
  \centering
  \subcaptionbox{Cohort 1.\label{fig:confidence-ndt}}[0.65\linewidth]{
    \includegraphics[width=\linewidth]{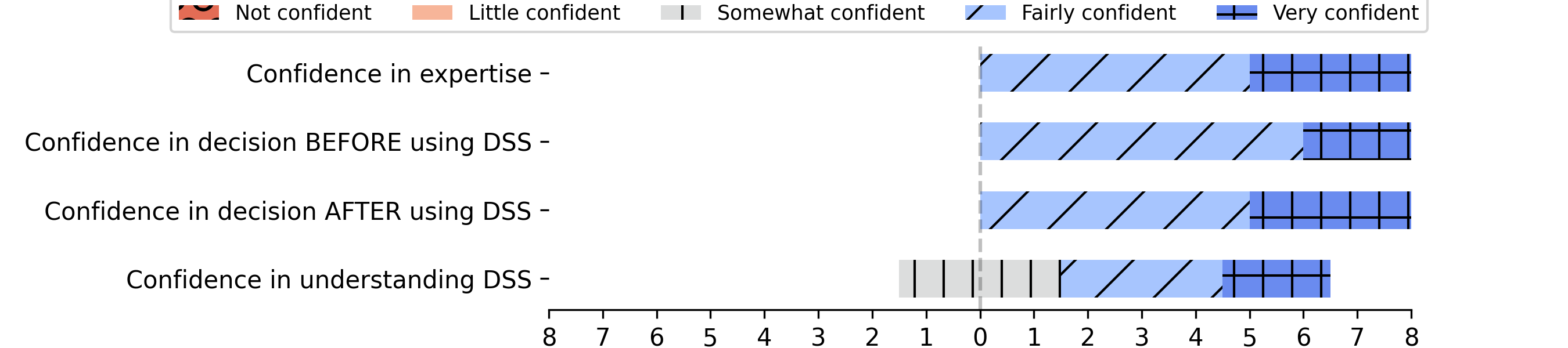}
  }
  \par
  \subcaptionbox{Cohort 2.\label{fig:confidence-endo}}[0.65\linewidth]{
    \includegraphics[width=\linewidth]{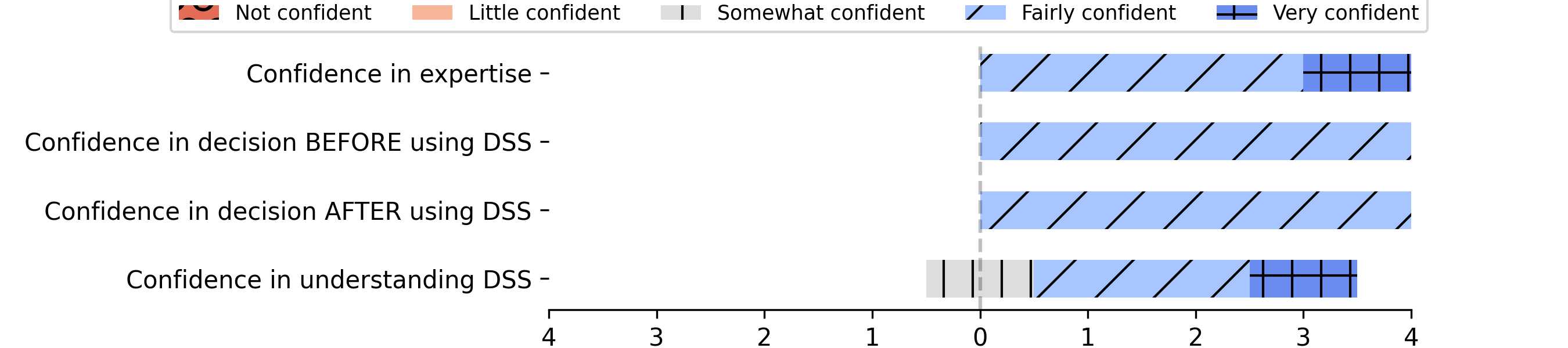}
  }
  \caption{Participants' confidence scores regarding their expertise and decision-making before and after using the DSS, as well as their understanding of the DSS.}
  \label{fig:confidence}
  \Description{Participants' confidence scores regarding their expertise and decision-making before and after using the DSS, as well as their understanding of the DSS.}
\end{figure}

In the interview, P1 suggested that, over the years, they acquired some tacit understanding of how the DSS works, ``we've been using [the DSS] for quite a long time. So that's a little bit too late to go into this kind of explaining [of] how the [DSS] works''. In cohort 2, the manual data entry helps participants to develop an understanding of how the DSS works. As P12 explained in the interview, ``So if you are using it [DSS] all the time and you can see from the model what it is based on, because you fill it in yourself, then you get some feeling for what it [prediction] means''. 
In view of this, participants rated their confidence in understanding the DSS as fairly confident (Fig.~\ref{fig:confidence}). Besides, participants rated their knowledge about artificial intelligence as slightly to moderately knowledgable (Fig.~\ref{fig:ai-knowledge}). %
Given their extensive experience as well as their (acquired) understanding of the DSS, the participants answered in the negative when asked whether they saw a risk in relying too heavily on the predictions.

\subsection{DSS Use Changes With Experience}
\label{sect:frequency-expertise}

Closely related to the above, some participants explained that the frequency of use has changed over the years as they gained more experience: ``So when I started, I used it [DSS] every time, with every patient. And now it's more when I feel, I don't know about the patient, then I start looking at it'' (P5). Similarly, P4 mentioned, ``If I look at myself for the last 10 years, in the beginning of my career, I looked at it [DSS] more than I do now''.
With more experience, the DSS became less important to them, except in more complex cases (section~\ref{sect:dss-doubt}).

One participant, however, reported the opposite. When asked how the use of the DSS had changed over the years, P10 referred to a different DSS, namely the ADNEX model for diagnosing ovarian cancer, which they have been working with for around six years. This is because the DSS in cohort 2 has only been in use since beginning of 2026. P10 mentioned that they gained more confidence in the DSS after using it for some time. Initially, they were more sceptical of the DSS since ``in the beginning, we didn't know those factors [of the DSS] as well''. Therefore, after the surgery, they validated the DSS prediction, ``we use it [the DSS] before surgery so we always have to check in the end if the model was right. And we kept on checking the model with the histology [study of tissue] for a long time''. Consequently, with time, they understood the workings of the DSS better, ``So in the beginning, I didn't rely as much on the model as I do now. I know better in which factors the model is not perfect. [...] So I know where the pitfalls are better after some years''. 
This procedure also highlights the critical attitude that prevails in clinical context (section~\ref{sect:critical-judgement}).

So while clinicians gained more confidence in their own expertise, which resulted in less DSS use, they also gained more confidence in working with and understanding the DSS, which increased DSS use. In either case, the use and reliance behaviour changed with experience.

\subsection{The Same DSS is Used for Different Purposes}

In both cohorts, the respective DSS is used in the same context, yet by different healthcare professionals, who use the DSS for (slightly) different purposes.
In cohort 1, the clinicians use the DSS for triaging patients, i.e., ``to get the right patient to the right doctor'' (P1), and for treatment selection. Similarly, in cohort 2, the DSS is used for 1) triage, 2) staging, i.e., determining the extent of the disease, 3) treatment selection, i.e., which type of surgery to perform, and 4) patient consultation.
For triage, the first decision point, the patient has not yet been admitted to the hospital, it thus has to be decided as to whether the patient should be admitted and, if so, to which specialist (e.g., surgeon or rehabilitation specialist) they should be referred. Once the patient is admitted, the clinicians use the DSS outcome to prepare for consultation, to inform the patient during consultation, and to make a treatment decision; unless the clinician decides during consultation that the patient should see another specialist (i.e., triage). 

Although these decision steps can overlap, some clinicians do primarily the triage (e.g., physician assistant), while others focus mainly on patient consultation and treatment selection (e.g., surgeon). 
This means that, in both cohorts, the same DSS is used at different stages of the patient's trajectory, and, crucially, by clinicians with different specialisations. Thus, the clinicians' information needs differ. As P7 explained in the open text field in the survey, ``I look at it from a surgeons perspective and I want to have an almost binary advice with nuances on complication and failure risks, and what percentage of improvement the operation will provide. From a perspective of rehabilitation or research the demands are different''.
Despite their different needs and objectives, our participants expressed similar views regarding the DSS, as described below.

\subsection{The DSS Predictions Make Decision-Making More Concrete}
\label{sect:objective-reflection}

\begin{figure}
    \centering
    \subcaptionbox{Cohort 1.\label{fig:use-ndt}}[0.8\linewidth]{
    \includegraphics[width=\linewidth]{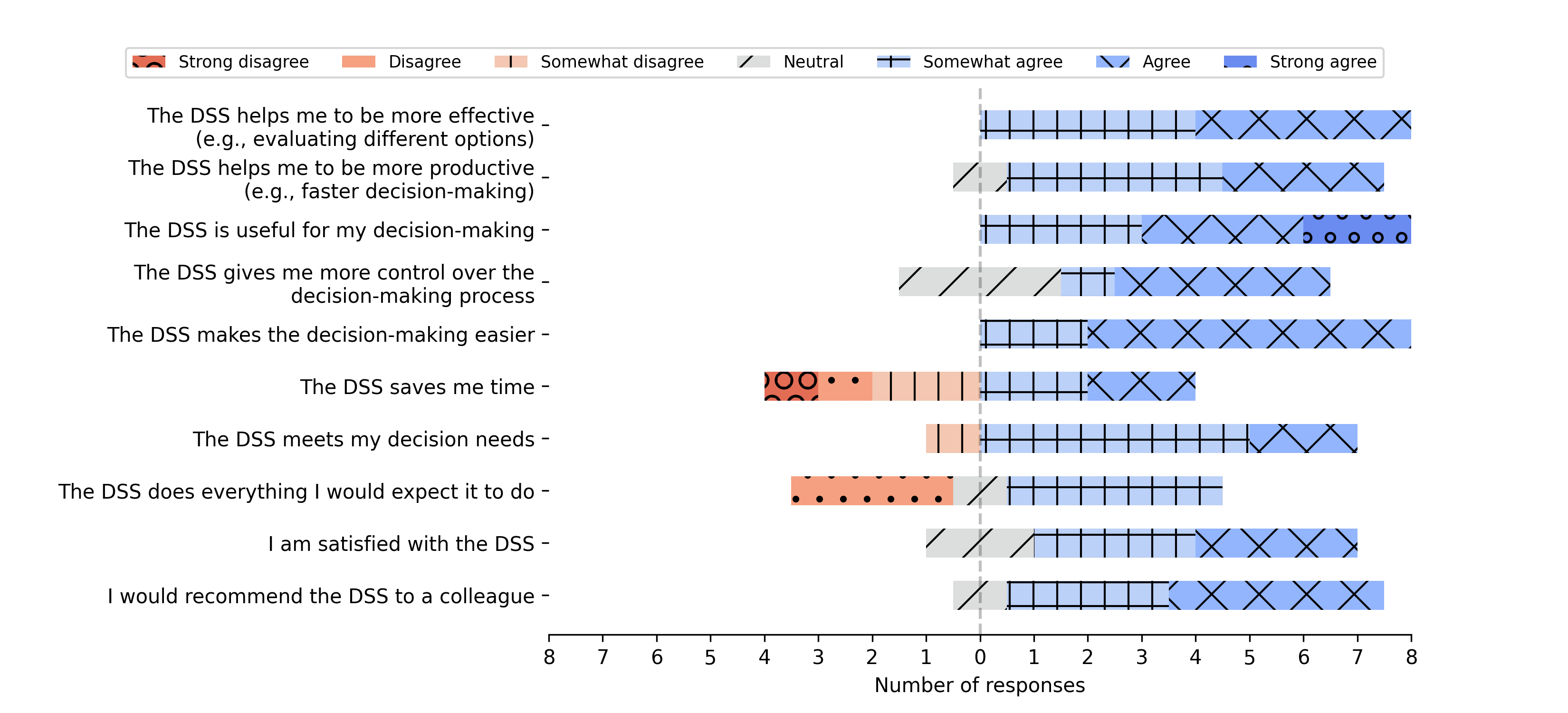}
  }
  \par
  \subcaptionbox{Cohort 2.\label{fig:use-endo}}[0.8\linewidth]{
    \includegraphics[width=\linewidth]{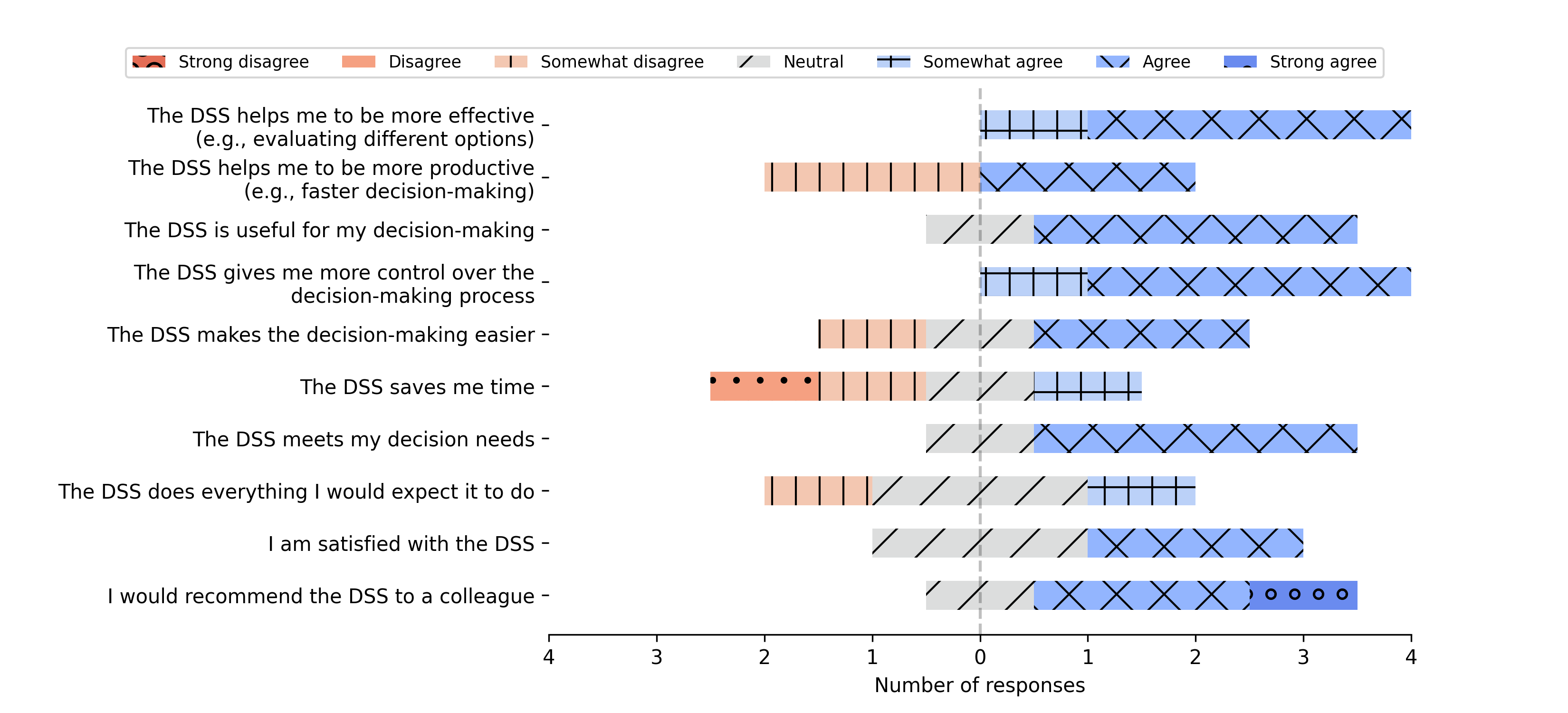}
  }
    \caption{Participants' attitudes towards the DSSs, as expressed by the Usefulness dimension of the Usability, Satisfaction, and Ease of Use (USE) Scale \cite{Lund2001}.}
    \label{fig:use-scale}
    \Description{Participants' attitudes towards the DSSs, as expressed by the Usefulness dimension of the Usability, Satisfaction, and Ease of Use (USE) Scale.}
\end{figure}

In each cohort, participants displayed rather positive attitudes towards the relevant DSS and rated it as `somewhat useful', with similar response patterns (Fig.~\ref{fig:use-scale}). 
The mean score on the Usefulness dimension of the Usability, Satisfaction, and Ease of Use (USE) Scale is 5 out of 7 (71.43\%) for both cohorts. For cohort 1, the median score is 5 ($SD=1.18$, $mode=5$), with the three highest rated items being, 1) ``The DSS makes the decision-making easier'' ($Md=6$, $SD=0.46$), 2) ``The DSS is useful for my decision-making'' ($Md=6$, $SD=0.83$), and 3) ``The DSS helps me to be more effective'' ($Md=5.5$, $SD=0.53$). For cohort 2, the median score is 6 ($SD=1.26$, $mode=6$), with the three highest items being, 1) ``The DSS helps me to be more effective'' ($Md=6$, $SD=0.5$), 2) ``The DSS gives me more control over the decision-making process'' ($Md=6$, $SD=0.5$), and 3) ``The DSS is useful for my decision-making'' ($Md=6$, $SD=1$). 
A reason for this positive attitude was provided in the open-ended questions of the survey (\textit{How does the DSS help you with these challenges?}), where participants noted that the DSS provides ``strong numbers'' (P5), whereby it ``objectifies predictions'' (P9). In the interview, P10 elaborated, ``the models are helpful because I think every doctor already calculates some kind of a risk, but not in real numbers, [...] so it's more concrete''. P10 went on to say that the number ``helps me to explain to patients how high the risk is''. 

\begin{figure}
    \centering
    \subcaptionbox{Cohort 1.\label{fig:interaction-ndt}}[0.8\linewidth]{
    \includegraphics[width=\linewidth]{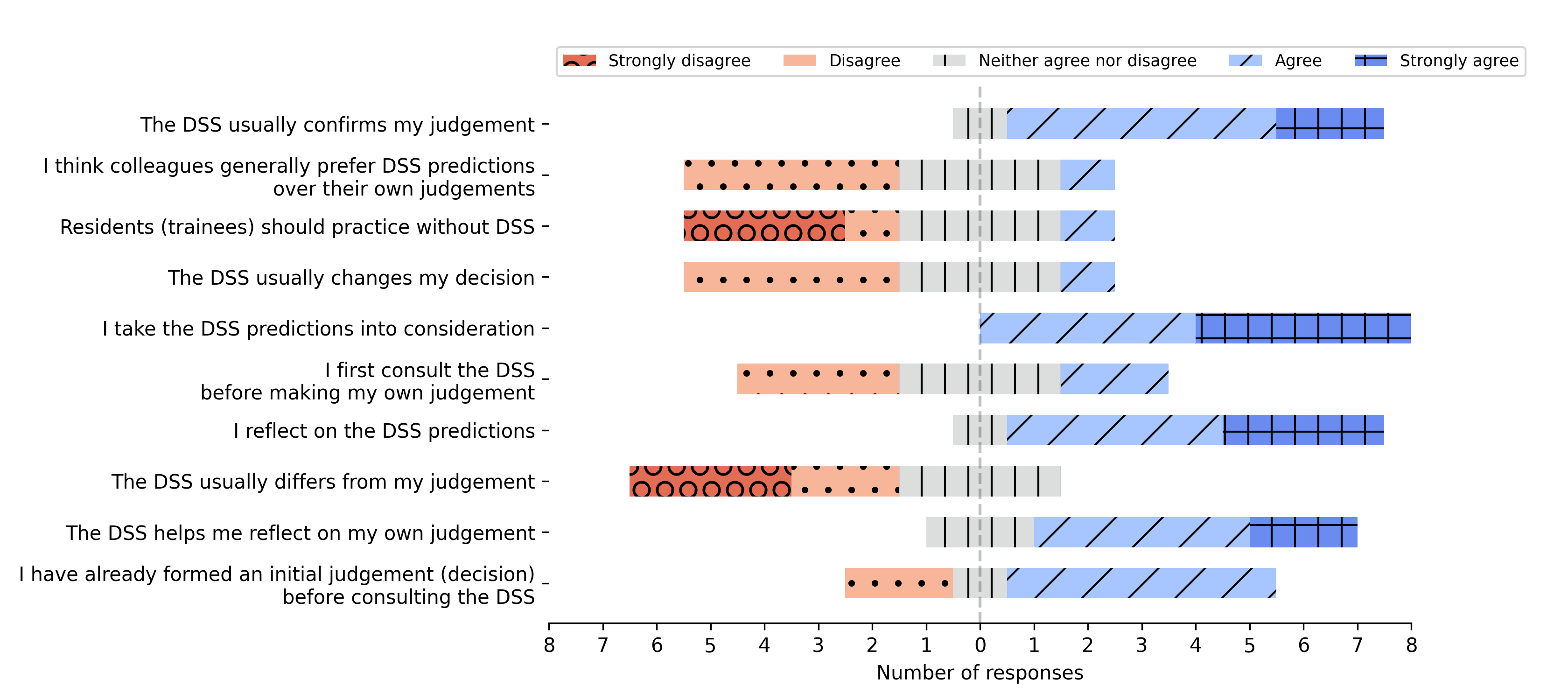}
  }
  \par
  \subcaptionbox{Cohort 2.\label{fig:interaction-endo}}[0.8\linewidth]{
    \includegraphics[width=\linewidth]{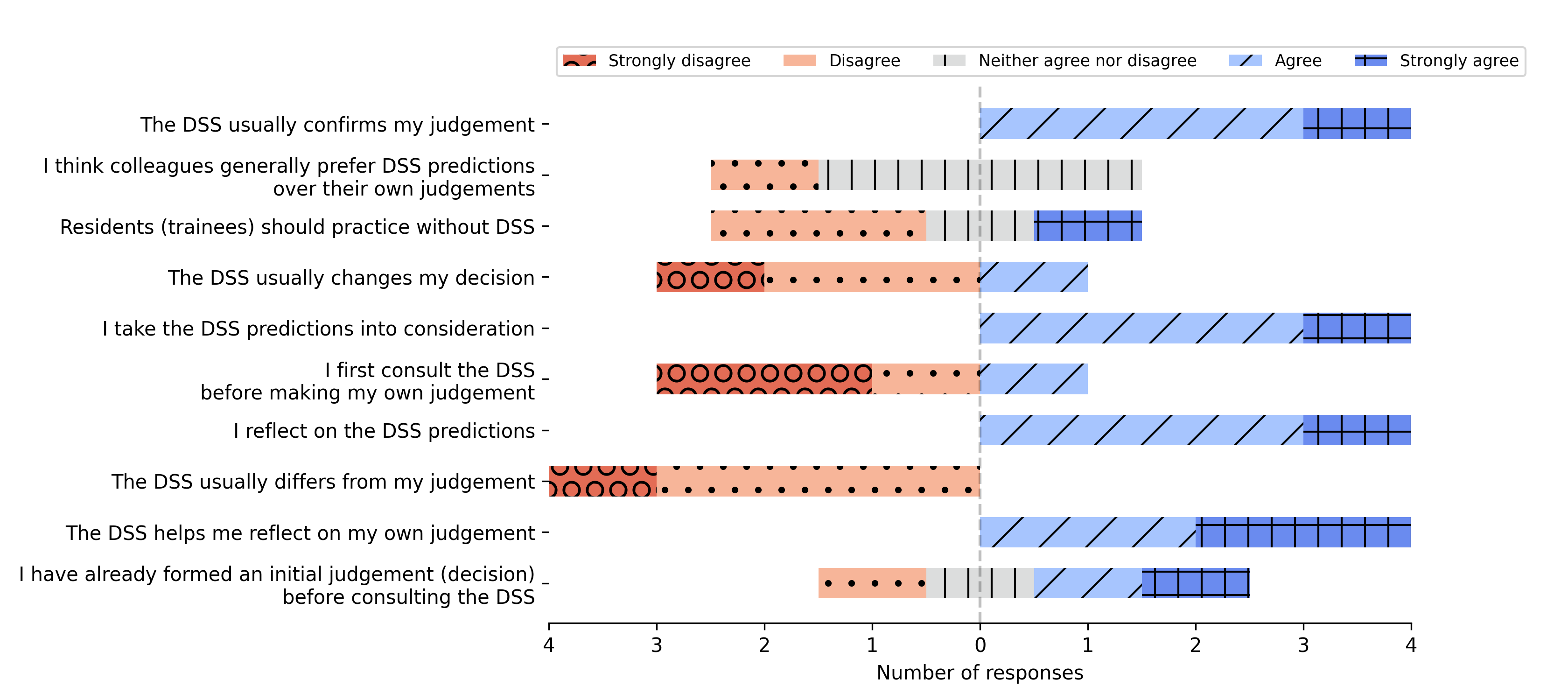}
  }
    \caption{A visualisation of the Likert scale regarding interaction behaviour and how clinicians integrate the DSS predictions into their decision-making process.}
    \label{fig:interaction-likert}
    \Description{A visualisation of the Likert scale regarding interaction behaviour and how clinicians integrate the DSS predictions into their decision-making process.}
\end{figure}

In view of the above and as expressed in the survey (Fig.~\ref{fig:interaction-likert}), all participants take the predictions into consideration when making a decision. While doing so, the majority of participants have already formed a judgement before consulting the DSS (Fig.~\ref{fig:interaction-likert}). P12 elaborated in the interview, ``since we do a standardized workup for every patient, we were already aware which of the patients would be high risk and which of the patients would be low risk, maybe not in these, um, hard percentages''.
As Fig.~\ref{fig:interaction-likert} further shows, the predictions generally confirm the clinician's assessment and rarely change their decision. The ``strong numbers'' (P5) therefore provide some reassurance.

One participant, however, pointed out the risk of a false sense of certainty or security. 
In the open question in the survey regarding challenges in decision-making, P12 mentioned the ``insecurity about the value of different items and the impression of security that the DSS provides''. This impression of security can lead to overreliance.

\subsection{A Critical Attitude Encourages Clinicians to Follow Their Own Judgement}
\label{sect:critical-judgement}

With regard to appropriate reliance, challenges arise when the decision-maker's judgement differs from the DSS predictions. P5 highlighted this challenge in the interview, ``Sometimes, it [the prediction] doesn't match at all to what you think. And then you should decide. What do you do? Do you believe the DSS, or do you believe your own judgement?''
In the open questions in the survey, most participants noted that in the case the DSS prediction contradicts their own assessment, they follow their own judgement. As Fig.~\ref{fig:confidence} illustrates, participants have slightly more confidence in their expertise than in their understanding of how the DSS works. Furthermore, in the interview, P1 stated, ``I'm not going to change my recommendation just based on the DSS. It's just a prognosis, so it's not fact''. A similar critical attitude was voiced by P2, saying that ``it's really more about reading between the lines than trusting completely on the [DSS]''. 

All participants indicated in the survey that they reflect on the DSS prediction (Fig.~\ref{fig:interaction-likert}). As such, they responded to the question \textit{What do you do if your judgement and the DSS prediction differ}, ``consider why it [prediction] can differ'' (P7) and ``try to explain why they differ'' (P9). 
P9 elaborated in the interview, ``In the hospital there is this type of attitude, like, question everything. So somebody says it's this, but is it really that, you know? So look at all the reports and pathology and radiology very critical. Is it really that or should we think of anything else?'' %
This questioning attitude was also expressed by participants from cohort 1 in the survey, indicating that they would like to know more about the quality and reliability of the data from the subjective patient questionnaires (Fig.~\ref{fig:taxo-ndt}). In addition, they indicated that they consider the consequences of a decision (Fig.~\ref{fig:tax-needs}). 
Three participants from cohort 1 mentioned in the interviews that their department tries to investigate the feasibility and effectiveness of other treatment options before opting for surgery.
In both cohorts, therefore, careful and critical decision-making processes take place, which are shaped by the (medical) context.

\subsection{The DSS Predictions Help Clinicians Reflect on Their Judgement}

Despite the clinicians' critical attitude and the DSS providing ``strong numbers'', which usually confirm their judgement, participants also mentioned in the interviews that the predictions affect their decision-making and might require them to justify certain reasons. As P1 explained, ``So if the [DSS] says this is a very bad plan to do surgery because the percentage of success rate is less than 10\%, then you must have a very good reason to do surgery. So it affects your decision-making to some extent''. P6 further explained that the predictions help to compare it with their own judgement, ``You have something in mind and this [prediction] really helps your decision. Like, if it matches your assumption, you think, okay, I can be confident, this can be the best decision. If it doesn't match, then it makes you reflect on your decision''. The predictions can thus help clinicians to reconsider and clarify their own judgements that they have formed without the DSS. 
Consequently, as P2 mentioned, ``does it [DSS] change my initial judgment? Sometimes it does''.

\begin{figure}
    \centering
    \subcaptionbox{Cohort 1.\label{fig:taxo-ndt}}[0.97\linewidth]{
    \includegraphics[width=\linewidth]{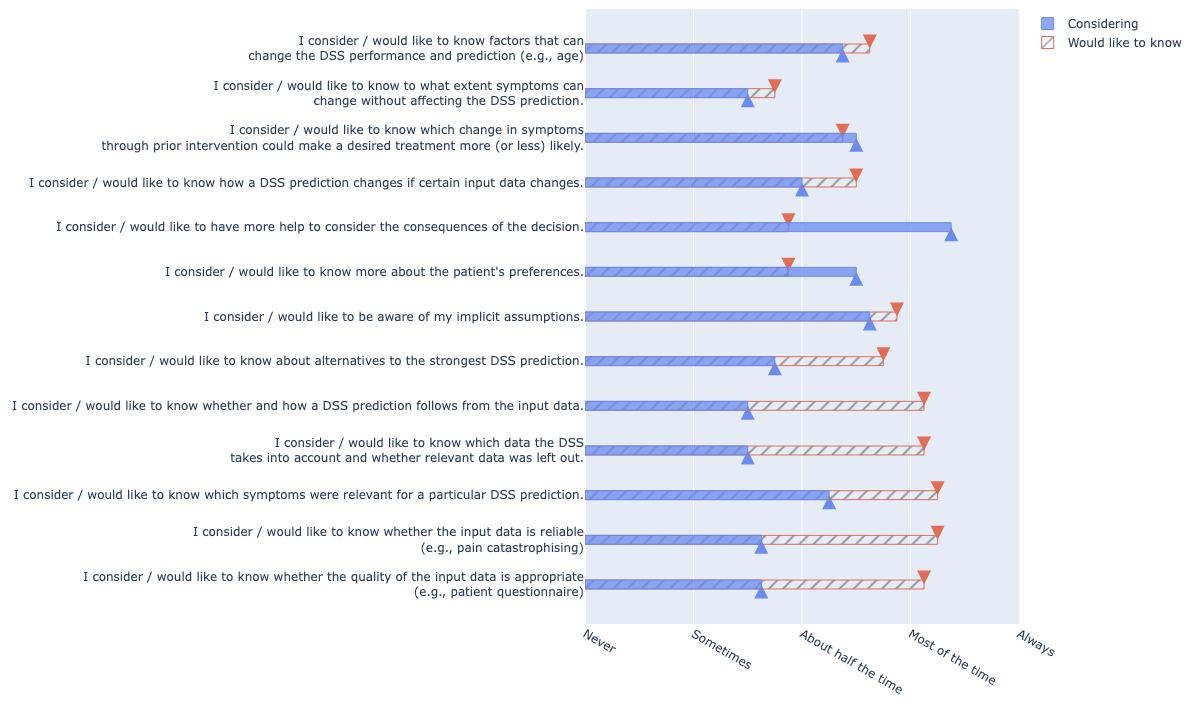}
  }
  \subcaptionbox{Cohort 2.\label{fig:taxo-endo}}[0.97\linewidth]{
    \includegraphics[width=\linewidth]{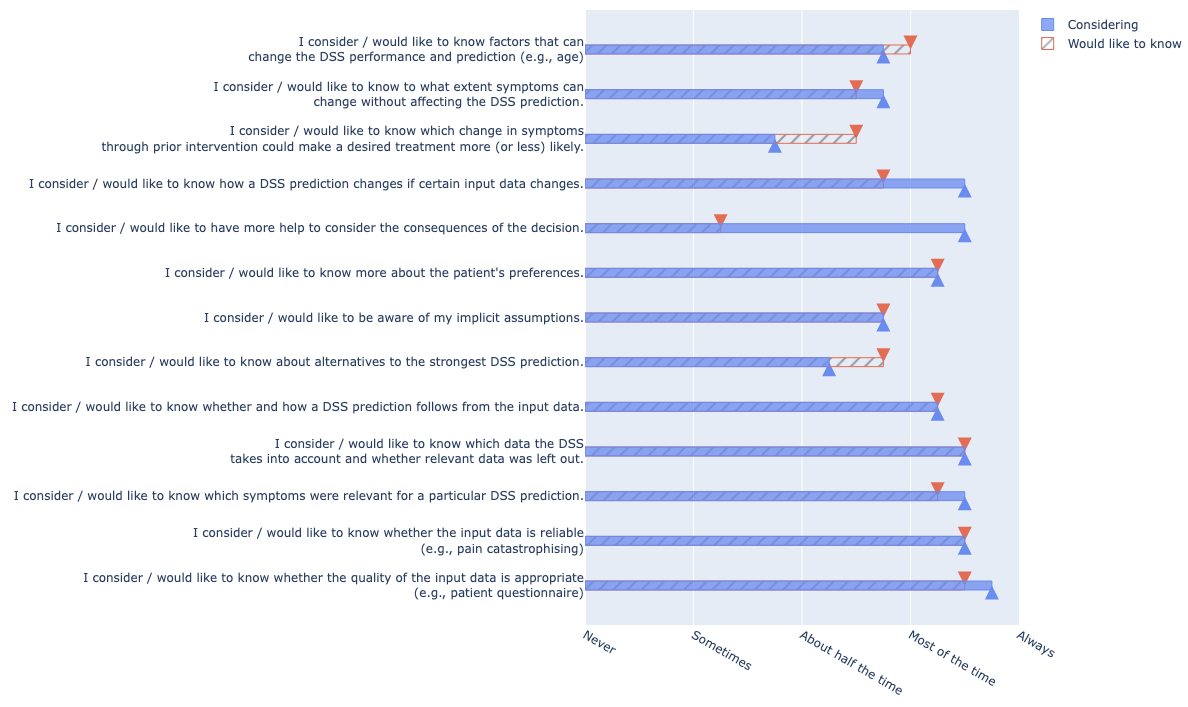}
  }
    \caption{Clinicians' informational needs, as determined by comparing the extent to which they consider certain information with the extent to which they would like to know more about that information.}
    \label{fig:tax-needs}
    \Description{A diagram showing the extent to which clinicians consider certain information and to what extent they would like to know more about that information. The difference in these two represent an informational need.}
\end{figure}

To better compare the prediction with their own judgement, P5 stated that a ``hard number'', such as feature contributions, would help. Similarly, P7 mentioned in the interview, ``It would help me if I would know which factors steer me against, or give a direction towards this [prediction], because then you can evaluate whether you're biased or not. I think that, for me, that would be important''.
As such, as Fig.~\ref{fig:taxo-ndt} shows, participants in cohort 1 expressed a need for explanations.
In cohort 2, the DSS already provides feature contributions, and accordingly, a need for explanations was not expressed (Fig.~\ref{fig:taxo-endo}). In the interviews, the participants confirmed that the provided explanations are helpful for their decision-making.

As Fig.~\ref{fig:tax-needs} further shows, participants in both cohorts indicated a need to investigate alternative outcomes to the prediction and to examine how changes to factors (i.e., input data) affect the outcome. P12 discussed the scenario in which a patient might be too weak for the suggested treatment, ``my patient is too vulnerable for the type of treatment that I'm proposing, but [then] I need to know [the] results of the alternative treatment''. Similarly, P4 expressed the need for a `what-if' analysis by explaining that prior surgeries performed years ago might be negligible, ``if you could see like a bar here, which [...] would sort of predict what it [prediction] would be if there were no prior surgeries, and I think that might help''.
Predicting alternative outcomes based on adjusted data could therefore help participants to further reflect on their assessment and, ideally, find the most appropriate treatment for a patient.

\subsection{The DSS Predictions Become More Important When Clinicians Are in Doubt}
\label{sect:dss-doubt}

The participants are fairly confident in their expertise and decision-making (Fig.~\ref{fig:confidence}). In complex cases, however, the predictions become more important and have a greater influence on the clinicians' reasoning, given that ``usually the extremes are the most easy ones'' (P1). Participants mentioned in the interviews that the DSS is especially helpful when they are in doubt. As P4 elaborated, ``If there is doubt about the surgery, we sometimes use [the DSS] to give us an idea, is this the right patient to do this more or less doubtful surgery, or not''. P2 added, ``When I'm in doubt, I consider the [DSS]. If it's a really clear case, I don't even look at the [DSS]''.
Similarly, P3 noted, ``I only use it [the DSS] when I’m a bit in doubt. And then when I see really good numbers [predictions], you think, okay''. 
In view of this, the DSS predictions can function as a second opinion, which can help to decide. P6 explained, ``sometimes I rather follow the [DSS]. And I say, oh yeah, I didn't take that [factor] into account''.
As a result, the DSS predictions can increase the clinician's confidence in the decision.
As Fig.~\ref{fig:confidence-ndt} shows, one participant indicated an increase in their confidence in the decision \textit{after} using the DSS compared with \textit{before} using the DSS.

\subsection{Inexperienced Clinicians May Be More Influenced by DSS Predictions}
\label{sect:inexperienced}

Given that our experienced participants consider the DSS predictions more when they themselves are in doubt, (section~\ref{sect:dss-doubt}), they expected inexperienced decision-makers to be more prone to overreliance on the DSS. When asked whether less trained colleagues might rely more on the predictions, P2 replied ``Yes. When they're really in doubt, I think they rely sometimes too much on the [DSS] scores''. This doubt can stem from a lack of experience, as P10 pointed out: ``They don't have the calculations in their head yet because they don't have the experience yet. So they probably rely more on the model for measuring the risk''.

When asked if it is necessary to stimulate critical thinking in clinicians, P9 replied ``I think that's a long term issue. Because now we are working with people that have like multiple years of clinical experience without these kinds of models. So they are being trained properly, but in five years, there are new people''.
This hints at the importance of skill development and the need to train young clinicians in such a way that they can make judgements without a DSS. Nevertheless, most participants indicated in the survey that residents (trainees) should practice with the DSS (Fig.~\ref{fig:interaction-likert}). 
Participants in cohort 2 mentioned that clinicians receive a training on how to use the DSS. P9 noted, ``we have some training before you use the model''. Similarly, P10 mentioned, ``they [inexperienced clinicians] really have to do a course before they can use the model. So they do have a day of training and explanation''. %
While this training helps inexperienced clinicians to become more familiar with the DSS, their lack of professional experience may nevertheless make them more susceptible to ``the impression of security that the DSS provides'' (P12).

\subsection{The DSS Predictions Are Just One Aspect of a Multi-Faceted Decision-Making Process}
\label{sect:multi-faceted-process}

The clinical decision-making process consists of multiple steps and various sources of information. 
For the open question in the survey (\textit{What are the challenges you face in your decision-making process?}), P9 replied ``A lot of factors contribute to decision making (patient factors, comorbidity, preferences, need for additional treatments, toxicity of treatment)''.
Similarly, P4 explained that depending on additional data, like magnetic resonance imaging (MRI), the DSS predictions become more or less important, ``So, if you have an MRI scan, which has a clear surgical indication, the DSS is not really relevant. When the MRI scan says, difficult, you might do surgery, don't have to, then it [the DSS] becomes increasingly important''. 
P5 corroborated the importance of additional information, ``Even if the [prediction] scores are horrible, but I see the patient's story is exactly what you should expect in the imaging or in the physical examination, then even though the decision tool says differently, then it feels a little bit weird to only trust on the AI''. P1 mentioned something similar, ``And sometimes, despite the fact that the [DSS] is telling me, [...] the prognosis for success of surgery is 25\%, and still, because of your clinical exam and your diagnostic test, you say, okay, still we're going to recommend the surgery, and the patient may do very well after the surgery''.
P7 explained that the DSS itself, although useful, is not sufficient to make a decision because ``you need the context of the patient, you need to know what the medical history is''.
Moreover, in the survey, P11 and P12 pointed out the importance of shared decision-making with the patient. P12 wrote for the open question asking about what they do when the prediction contradicts their judgement, ``[I] try to find an explanation and weigh that with the patient and her preference''.
In view of the above, the DSS only provides one piece of information in the decision-making process, which must be weighed up accordingly.

This raises the question what data to include in the DSS.
The surgeons in cohort 1 mentioned that they would benefit if the DSS made use of additional diagnostic data, such as magnetic resonance imaging (MRI). Although these MRI scans exist for most patients, as they have been to previous specialists, the current DSS does not take this data into account. As P1 explained, ``there's still no link to additional diagnostics. So we cannot correlate this [predictions] to MRI results''. Consequently, the surgeons have to make an informed decision by piecing together different sources of information, e.g., DSS predictions and MRI scans. For this reason, participants in cohort 1 expressed dissatisfaction with the DSS on the USE Scale (Fig.~\ref{fig:use-ndt}), with ``The DSS does everything I would expect it to do'' ($Md=4.5$, $SD=1.49$), and ``The DSS meets my decision needs'' ($Md=5$, $SD=0.93$), as the second and third lowest rated items respectively. 
Ideally, the DSS takes into account these different sources of information to form a coherent picture of the patient. However, in a text-field for open remarks in the survey, P7 noted, ``There is always a cut off between the ideal complete version and the most user friendly version. Finding the right balance is in my opinion the biggest challenge''.

\subsection{Decisions are Discussed with Colleagues}

With regard to the multifaceted decision-making process, our participants noted that they discuss cases with colleagues to varying degrees. For complicated cases, P3 mentioned, ``Sometimes I go to the colleague and I say, okay, we can do surgery, but it's a really, really low score on the [DSS]. And then we go, why?''.
Furthermore, in both cohorts, all cases are discussed in weekly group meetings. For cohort 1, P6 explained, ``Well, every Thursday evening, we have a discussion of patients in the spine unit. And if you hesitate you can put them on the list. And then we discuss with eight or ten [colleagues] and take everything into account. Like the radiology, but also the [DSS], also the [patient's] story. It's good to have the opinion of your colleagues''. The outcome of these meetings could result in having the patient come to the clinic once again. As P6 explained, ``And sometimes we just say maybe it's good to let him [the patient] come another time to one of the colleagues. To have an internal second opinion, let's say, because you have black and white, but you have a lot of grey [\textit{sic}]''. Accordingly, P4 replied for the open question in the survey (\textit{What do you do if your judgement and the DSS prediction differ?}), ``I will discuss it in the spine group''. 
In cohort 2 similar group meetings take place. In the interview, P9 elaborated, ``So we have this multidisciplinary tumour board with like the gynaecologist, pathologist, like the whole treatment team. And it's also with the gynaecologist from the smaller hospitals. Then we already give some treatment advice, like you should do this type of surgery or this type of surgery. And now we also use the [DSS] in the multidisciplinary team as well. It's like a quality assurance thing. All oncological diagnoses have to be discussed''. 

In addition to these weekly group meetings, in cohort 1 all the surgical cases are discussed amongst the surgeons one week before the scheduled date. In this way, different opinions on complex cases can be sought and discussed, which, in some (rare) circumstances, may lead the surgeons to cancel the surgery. P5 elaborated, ``a week before the surgeries we get together, we look at all the surgeries we're going to do. So, that's also a moment, where we could debate with the group, are you sure about this?''
Consequently, these meetings represent an additional decision point, after triage, patient consultation, and weekly group meetings.

These group meetings are important for at least two reasons. First, inexperienced clinicians also take part in these discussions. 
This enables them not only to discuss cases where they are uncertain, but also learn to be more critical towards the DSS outputs  (section~\ref{sect:inexperienced}). The social aspect of decision-making and learning was highlighted by one participant during the interview. When asked how inexperienced clinicians could be helped to remain critical of the DSS, P12 replied ``I would say in general that would be helped by something like a tumor board because you're always working in a clinic with more senior physicians, more junior physicians. So the juniors get to learn from the seniors, uh, and the seniors get to learn from the juniors as well''.
Second, the discussions amongst colleagues highlight the importance of (medical) consensus. P12 pointed out, ``we haven't decided yet, we meaning the professional community, what the best cut-off [threshold value] is in the model''. So whether, for example, surgery should be performed if the risk score is 10\%, or rather, 15\%. Hence, on the one hand, standards or guidelines through consensus are important to guide the decision-making process. On the other hand, consensus can contribute to a shared understanding of the DSS, similar to the case described above where the prediction was compared with the histology (section~\ref{sect:frequency-expertise}), and help to raise awareness of shortcomings, which could ultimately help to improve the DSS.

\subsection{Workflow Integration of the DSS Affects Decision-Making}

For cohort 1, the DSS is (or rather was) integrated into the electronic patient file. Most participants, however, mentioned complications regarding workflow integration. The hospital's digital infrastructure or medical record system has undergone various changes and updates over the years, resulting in difficulties using the DSS at times. As P1 elaborated, ``Because at first we had Hix 6.1, now we have Hix 6.3, but in between there was another company that probably could take care of the electronic patient file. But then that turned out to be not really matching to the way we want to run this hospital and what we would like to see in the electronic patient file. So then the contract was cancelled. Now we're back to Hix, but Hix as it is today is not Hix as it was before. It's completely different and it's more difficult to trace all these data''. 
As the software provider (i.e., Hix) had updated their product, the clinicians encountered interoperability issues between the systems, i.e., electronic patient file and DSS, which ultimately meant additional software had to be introduced. As P2 explained, ``Previously, we had links to Hix, so we could all open it [patient file] just when we were preparing our outpatient clinic, and we could just open the [questionnaire] response of the patient and this conclusion [prediction]. But since we went to the Hix 6.3, the link is gone. So, the [DSS predictions] and all the questionnaires, we can't open them any more in Hix. So, we have to work with the Power BI app, but therefore we have to buy some licenses to log in''. The additional software component (i.e., Power BI app), and its licences added accessibility and functionality constraints. Now, only the two or three clinicians who do the triage have direct access to the DSS and its predictions (Power BI), which is why they have to leave notes for their colleagues in the patient record (Hix). Accordingly, this requires more time, as P3 noted, ``That costs us a lot of more work. It's not in Hix any more, it's now in Power BI. So we have to note all the scores''. 
The daily use of the DSS is constrained by the digital infrastructure of the hospital, which relies on other providers. Consequently, as P4 put it, ``it's a brilliant system [DSS], but if it goes on and off of [stops] working, then you tend to not use it [DSS] so much''.

For cohort 2, the DSS is not integrated into the electronic patient file. Instead, as P9 explained, clinicians need to manually enter the patient data into the DSS. Accordingly, the statements with the lowest and third lowest rating on the USE Scale (Fig.~\ref{fig:use-endo}) are ``The DSS saves me time'' ($Md=3.5$, $SD=1.29$), and ``The DSS helps me to be more productive'' ($Md=4.5$, $SD=1.73$). 
Nevertheless, the fact that the clinicians need to enter and engage with the data means they have already formed an initial judgement before seeing the predictions (Fig.~\ref{fig:interaction-endo}). Clinicians from cohort 1, however, also reported that they have already formed an initial judgement before consulting the DSS, although the predictions are available to them upon opening the patient file (Fig.~\ref{fig:interaction-ndt}).

\section{Discussion}
\label{sect:discussion}

\begin{figure}
    \centering
    \includegraphics[width=0.85\textwidth]{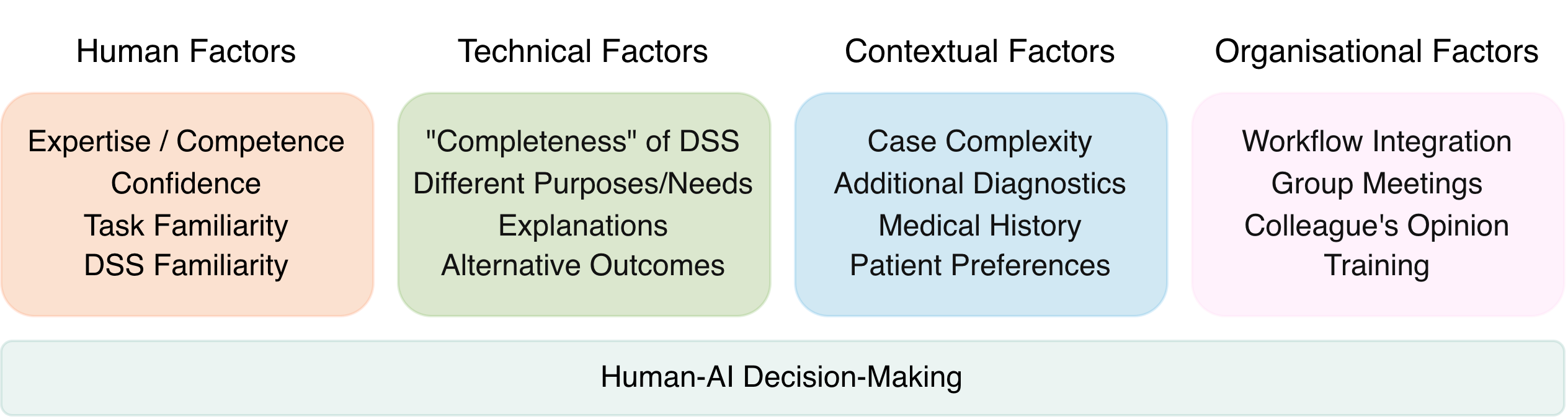}
    \caption{Socio-technical factors that shape the human-AI decision-making process in clinical practice, which we have identified on the basis of our results.}
    \Description{Socio-technical factors that shape the human-AI decision-making process in clinical practice, which we have identified on the basis of our results. They consist of human factors, technical factors, contextual factors, and organisational factors.}
    \label{fig:factors}
\end{figure}

We presented insights into how clinicians work with a decision support system to make decisions in clinical practice.
The eleven themes we identified form a socio-technical perspective on human-AI decision-making. Compared to \citet{Eckhardt2026}, who identify three aspects, namely the human, technical, and the environment (e.g., an organisation), we add a fourth factor (Fig.~\ref{fig:factors}), resulting in: 1) human factors, such as confidence of the decision-maker, 2) technical factors of the DSS, such as its completeness, 3) contextual factors such as additional diagnostics, and 4) organisational or workplace factors, such as group meetings, which \citet{Eckhardt2026} describe as the surrounding environment. 
Importantly, these four dimensions are not independent of each other but instead interact with one another and together shape the human-AI decision-making process \cite{Eckhardt2026}.
Based on these findings, we want to discuss some implications for the design of both DSS and approaches for promoting appropriate reliance on DSS.

\subsection{Human Factors}

With regard to the \textit{human factors}, the many years of experience of our participants highlight the importance of expertise and familiarity with both the task and the DSS for appropriate reliance. This complements work by \citet{Dogru2025}, who found that expertise was the most influential aspect for reliance behaviour, as experts are better at ``distinguishing between accurate and inaccurate AI advice'' and ``they are also less prone to rely on their own incorrect decision when confronted with correct AI advice''. Similarly, our participants mentioned that the discrepancy between their own assessment and the DSS prediction functions as a form of friction that allows them to reflect on this difference. 
In this, intuition, which is linked to expertise, also plays a role, as \citet{Chen2023b} point out.

Although decision-makers are often distinguished between experts and novices or laypeople in various studies \cite{Raees2026b}, it is rarely considered how they gain experience, especially with a DSS through prolonged use. Our participants mentioned that the frequency of DSS use changed throughout their careers. As they gained experience, their confidence grew, both in terms of their own decision-making and their understanding of the DSS. So while factors like task familiarity, AI familiarity or affinity for technology are discussed in the literature \cite{DeSouzaFilho2026,Kuper2025,Dogru2025}, less (or no) attention is given to the familiarity with the DSS. This might be partially due to the fact that many studies focus on single interaction sessions with novel prototypes. At the same time, there is the (technical) question of how a DSS and approaches for appropriate reliance can be designed in such a way that they remain useful to decision-makers even as they become more familiar with the decision task and with the DSS. 
In view of this, \textbf{the long-term use of a DSS should be given greater attention}.

\subsection{Technical Factors}

For the \textit{technical factors}, our results show that the DSS confirms our participants' judgements in most cases and rarely changes their decision. This raises the question as to what the DSS provides, besides reassurance through strong numbers. In complex and uncertain cases, this `second opinion' provided by the DSS can help to decide.
However, the fact that the DSS increases their confidence in the decision also increases the risk of overreliance \cite{Gaube2021}, especially in those complex cases.

In this regard, although not the primary focus of our research, our findings suggest that \textbf{the value of explanations remains unclear}. All participants rated their confidence in understanding the DSS (global level) as similarly high, even though the DSS in cohort 1 does not show explanations, while the one in cohort 2 does. Besides, participant P12, who expressed concerns regarding a false sense of security that might arise from the predictions, belongs to cohort 2, where the DSS provides explanations (local level). This also hints at the danger of explanations increasing the system's trustworthiness and thus increasing overreliance, as reported by others \cite{Schemmer2023,Jacobs2021a}.
Consequently, as suggested by \citet{Miller2023}, \textbf{it might be more beneficial to support decision-makers in ways other than providing recommendations or predictions} accompanied with explanations. One example, as expressed by participants and that is also reported by \citet{Liao2020}, could be the possibility to investigate alternative treatments through counterfactuals and hypothetical scenarios. %
This could potentially also promote discussions with colleagues \cite{Bach2023}. 
In addition, DSS are often portrayed as systems with a single objective, e.g, treatment prediction or diagnosis. In both cohorts, however, different clinicians with different areas of specialisation use the same DSS for different purposes and at different stages of the patient's trajectory, i.e., triage and consultation. Consequently, when focusing on the interplay of the human and technical factors, \textbf{more attention should be paid to diverse needs within the same context}.

\subsection{Contextual Factors}

In view of the \textit{contextual factors}, our findings show that the decision-making process in both cohorts is multifaceted and consists of multiple sources of information. \citet{Matsumoto2025} also emphasise the dynamic and complex nature of clinical decision-making. As such, it could be that other medical diagnostics, such as MRI scans, invalidate the DSS predictions and/or explanations. Similarly, patient preferences might result in a different decision despite an accurate prediction. 
\citet{Eckhardt2026} describe this as ``multi-factorial effects'', where multiple factors, such as predictions, explanations, and other contextual factors, interact.
In view of this, and given our participants expressed the need for explanations, it could be that explanations prove helpful in mitigating overreliance, although different studies show inconclusive and even negative effects of explanations \cite{Vasconcelos2023,Bansal2021}. 

Furthermore, our two use cases suggest that critical and reflective decision-making processes take place in a clinical context. In combination with the human factors (i.e., expertise), it thus appears that overreliance on DSS in clinical practice might be less of a problem than portrayed in the research literature, which often focuses on novel, unfamiliar systems in single interactions \cite{Gaube2021,Jacobs2021a,Raees2026b}. Overreliance nevertheless remains an issue, as our participants also recognised, particularly among less trained people.

It is therefore all the more important, \textbf{when evaluating strategies for appropriate reliance, to move from simplified decision-tasks to realistic use cases}.

\subsection{Organisational Factors}

Lastly, our findings highlight the importance of the \textit{organisational factors} in the decision-making process.
Both cohorts have a good governance structure, which include group meetings in which cases are discussed, collegial support in the form of a second opinion, and training sessions of how to use the DSS. All of these aspects help clinicians to learn from one another and develop decision-making skills \cite{Matsumoto2025}, which can help to promote appropriate reliance on DSS. To develop necessary skills, \citet{Ferdman2026} therefore argues that \textbf{``environments must encourage (not merely provide) intersubjective, embodied interactions to compensate for AI-mediated, disembodied environments''.} 

The collaborative decision-making process is also highlighted by the multiple decision steps, from triage through consultation with the patient to the group discussions. At each stage, it is likely that different clinicians review the same patient case and thus reach a decision together, or rather one decision impacts the next \cite{Matsumoto2025}.
So although the interaction with the DSS might be on an individual level, \textbf{interactions should be designed to support teams in discussions and shared reasoning}. This could be encouraged, for example, by investigating alternative scenarios that prompt clinicians to articulate their reasoning, as also suggested by \citet{Bach2023}.

An additional reason to focus on realistic use cases and to \textbf{investigate DSS in existing workflows, instead in isolation}, is the interplay between the technical and organisational factors. 
All participants from cohort 1 we spoke to mentioned several changes in the hospital's digital infrastructure which caused disruptions in working with the DSS. For cohort 2, on the other hand, the DSS is not connected with the electronic patient record, so they must enter the data manually into the DSS. Although this interaction might be a disadvantage from an efficiency perspective, it can be beneficial for the decision-maker's cognitive engagement. Both \citet{Bucinca2021} and \citet{Fogliato2022} report lower AI reliance when decision-makers have to make a preliminary decision before seeing the DSS output. \citet{Eckhardt2026} claim that this two-stage decision-making approach is ``very unlikely in real-world applications'', yet our presented use case provides a counterexample. 
In any case, as others have argued, the lack of evaluations in existing workflows leads to uncertainty regarding the efficacy of DSS and ultimately their impact on patient outcomes \cite{Goldenberg2026,Cabitza2026,VanCalster2023}.

\section{Limitations and Future Work}
\label{sect:limitations}

We acknowledge some limitations of our research. Our sample size of $n=12$ provides context-specific insights and our qualitative research does therefore not aim to provide general conclusions on human-AI decision-making in clinical practice. The recurring themes we have identified in two different clinical contexts nevertheless highlight commonalities and thus the transferability of our findings. 
Although we could recruit only four participants from cohort 2, the response patterns in both the survey and the interviews were very similar to those of cohort 1. The consistent responses across both cohorts ultimately led to thematic saturation. Furthermore, our data stem from real-world decision-making contexts and from two different sources, i.e., survey and interviews. However, we draw from the clinicians' subjective experiences. Participants might say they follow their own judgement, yet it is uncertain to what extent the predictions have influenced them for a particular decision. Similarly, participants mentioned they are critical of the outputs, yet in situations where attention is low or they face time pressure, automation bias and overreliance might still occur. 
Of course, there are always limitations when it comes to replicating a real decision-making task, e.g., simulating patient consultation. Nevertheless, additional (quantitative) studies would help to provide a broader understanding of how DSS are used in practice. To this end, reliable measurement methods are required that can assess possible overreliance on DSS in a multifaceted decision-making context.
Future work could also consider participants with different levels of expertise. While our participants are very experienced, inexperienced decision-makers, who might be less confident in their decision-making, and who in addition just learn to use the DSS, will most likely incorporate the advice differently \cite{Gaube2021}.

\section{Conclusion}
\label{sect:conclusion}

We provided insights into how clinicians work with a decision support system to make decisions in clinical practice. To this end, we conducted an online survey ($n=12$) and semi-structured interviews ($n=10$) with 12 clinicians in total from two different medical cohorts, who respectively work with two different DSS in clinical practice. By triangulating descriptive statistics for the quantitative data with a thematic analysis of the qualitative data, we identified eleven themes. 

Given their extensive experience, both with the decision task and in working with a DSS, our participants did not consider overreliance a problem for them. Instead, the predictions offer them a `second opinion' in the form of concrete numbers, which help them to make the decision-making process more explicit and to reflect on their own judgement. Particularly in more complex cases, where they are in doubt, the predictions become more important. In view of this, participants also assumed that novice clinicians, who lack the experience, are more prone to overreliance. The participants' own reliance behaviour changed over the years as they gained more experience. They also mentioned how changes to the digital infrastructure of the hospital affected their DSS use. Overall, their critical attitude, prevalent in the medical context, makes them wary of the DSS predictions and to follow their own judgement. The DSS predictions need to weighed up against other medical information, which is why participants considered the investigation of alternative outcomes (e.g., `what-if' predictions) to be potentially useful for their decision-making. Lastly, decisions are discussed with colleagues in group meetings. 

Our findings relating to human (e.g., confidence levels), technical (e.g., explanations), contextual (e.g., patient preferences), and organisational factors (e.g., group discussions) 
contribute to a socio-technical perspective on human-AI decision-making. In addition, we provide pathways for the design of both decision support systems and approaches for appropriate reliance. As such, the long-term use of DSS and the gained experience of decision-makers should be considered, as well as diverse needs within the same context. Moreover, DSS should facilitate the decision-making \textit{process} at both the individual and collective levels within teams.
Importantly, future research on human-AI decision-making and appropriate reliance should adopt a socio-technical perspective and focus more on realistic use cases rather than on simplified decision-making tasks.

\section*{Generative AI Usage Statement}
The authors used Grammarly (with AI features turned off) to polish some sentences. We also used DeepL Translator to adjust individual sentence structures by translating the English sentence into another language and then back into English. We used noScribe (version 0.6.2) to transcribe the audio recordings of our interviews, from which we have used quotes throughout the article. We have checked the transcripts for accuracy, as described in the methods section. Apart from that, no other generative AI tool was used in the creation of this paper.

\begin{acks}
\end{acks}

\bibliographystyle{ACM-Reference-Format}
\bibliography{Library}

\appendix

\end{document}